\documentclass[preprint]{aastex}

\usepackage[]{graphicx}

\begin{document}

\title{Old White Dwarf Stars with Some Hydrogen -- Cooling Curves}
\author{Eugene Y. Chen\altaffilmark{1} \& Brad M. S. Hansen\altaffilmark{2}}
\altaffiltext{1}{Department of Physics, The University of Texas at Austin,
1 University Station C1500, Austin, TX 78712; eyc@mail.utexas.edu}
\altaffiltext{2}{Department of Physics \& Astronomy, University of California Los Angeles, Los Angeles, CA 90095; hansen@astro.ucla.edu}

\lefthead{Chen \& Hansen}
\righthead{White Dwarf Stars with Some Hydrogen--Cooling Curves}

\begin{abstract}
We present theoretical analysis on old white dwarf stars with \emph{some} hydrogen, that possess a mass of surface hydrogen from $1\times10^{-11}M_{\sun}$ to $1\times10^{-7}M_{\sun}$.  The evolution of such objects is complicated by convective mixing from surface convection zone to the underlying helium layer.  In this paper, we provide first self-consistent, quantitative investigation on the subject of convective mixing.  Numerical cooling curves and chemical evolution curves are obtained as a function of white dwarf mass and hydrogen content.  Such results will be applied to the investigation of the non-DA gap of \citet{1997ApJS..108..339B} in a later paper.
\end{abstract}


\section{Introduction}
\label{sec:intro}
White dwarf stars are the burnt-out relics of lower mass main sequence stars and are by far the most abundant stellar remnants in the universe.  It is believed that up to 97\% of the stars in our galaxy will end up as white dwarfs (e.g. \citet{2001PASP..113..409F}).  As such, studies of white
dwarf populations provide invaluable information of the history of star formation and evolution in our galaxy. The modeling of white dwarf populations rests on the physics of white dwarf cooling, requiring a detailed treatment of matter under a wide range of pressure and temperature.
Most of the heat capacity resides in the dense carbon/oxygen core.  The rate of energy loss, on the other hand, is controlled by the thin non-degenerate envelope composed of hydrogen or helium.

The gravitational acceleration in white dwarf atmosphere is very strong, at the order of $10^8m/s^2$.  The gravitational field induces a non-vanishing electric field that separates chemical species of different charge to mass ratio.  In the absence of convective instability, the outermost part of a white dwarf atmosphere has a simplified structure consisting of an almost pure hydrogen layer on top of an almost pure helium layer \citep{1990ARA&A..28..139D,2001PASP..113..409F,2004PhR...399....1H}.

The rate of white dwarf cooling is significantly affected by the surface composition (e.g. \citet{1999ApJ...520..680H}).  It is most straight-forward to assume that the
composition of the white dwarf photosphere is determined by what remains after the mass loss on the Red Giant and Asymptotic Giant Branches, i.e., DB and DC stars are different from DA by lacking hydrogen.  However, observations suggest that the story may be more complicated.

It is observed that the spectral composition and effective temperature are correlated.  Two intervals in $T_{eff}$ space have been found to be almost devoid of non-DA stars.  Various mechanisms of spectral evolution have been explored and discussed qualitatively in the literature (e.g. \citet{2001ApJS..133..413B}).  The first gap resides at $40000K$ to $35000K$ and is interpreted to be the result of diffusion completion, followed by convective mixing from the underlying helium convection zone \citep{1987fbs..conf..319F}.  The explanation of the other gap (a.k.a. the ``non-DA gap''), located at $5000K$ to $6000K$, remains rather unclear \citep{1997ApJS..108..339B,2001ApJS..133..413B}.  Each of these gaps indicates that the photosphere composition of an
individual star could be variable during the course of white dwarf cooling.  The mainstream opinion is that, since a white dwarf can do nothing but cool, the two gaps suggests that a white dwarf could change its surface chemical composition from helium to hydrogen to helium to hydrogen to helium as it cools, due to the interplay of accretion of interstellar hydrogen and convective mixing.  However, such statement lacks quantitative verification.

In this series of paper, our goal is to \begin{enumerate}
\item Quantitatively investigate the interplay between spectral evolution and white dwarf cooling.
\item Explain the existence of non-DA gap and explore its cosmological implications.
\end{enumerate}

In the present paper, we hold the opinion that convective mixing (specifically, convective mixing from surface hydrogen layer to the underlying helium layer) is responsible for spectral evolution.  Our statement will be quantitatively validated.  We restrict ourselves to consider only the simplest scenario of convective mixing without any accretion from the interstellar medium (i.e., a model where total hydrogen mass, $m_H$, is fixed.)  It will be shown that the cooling curves produced from such scenario are already consistent with observations.  In principle, accretion can add a further dimension to the problem, but is unnecessary to capture the basic features of observation.

The paper is outlined as follows: In \S~\ref{sec:Mix} we briefly review past investigations on convective mixing, which is rather controversial.  In \S~\ref{sec:nomix} we study white dwarf envelopes in which convective mixing is absent.  We then introduce our self-consistent treatment of convective mixing in \S~\ref{sec:atmosphere} and revisit the convective mixing scenario in a quantitative fashion in \S~\ref{sec:sequence}.  In section \S~\ref{sec:evolve} we incorporate our results with the white dwarf evolution code of \citet{1999ApJ...520..680H} to calculate numerical cooling curves and chemical evolution curves.  Finally, we discuss and conclude our results in \S~\ref{sec:discuss} and \S~\ref{sec:conclude}.

\section{Brief Review on Convective Mixing} \label{sec:Mix}
Here, we specifically refer convective mixing to that from the surface hydrogen convection zone to the underlying helium layer.  Such scenario was first proposed by \citet{1972ApJ...177..723S} and was further investigated by \citet{1973A&A....27..307B} and \citet{1976A&A....52..415K}.  It can be shown that, at most observable $T_{eff}$, the surface hydrogen convection zone grows as $T_{eff}$ decreases. (see e.g., fig.~\ref{plottwo}) Thus, in the proposed scenario, a white dwarf with an intermediate amount of hydrogen (i.e., optically thick while less than $O(10^{-7}M_{\sun})$) will appear as a DA white dwarf \emph{until} the base of the surface convection zone reaches the underlying helium layer.  When the base of the convection zone reaches the underlying helium layer, Shipman claimed that the convective motion will dredge-up the underlying helium to the surface, resulting in a DB (helium lines) \emph{or} DC (no lines) white dwarf.  Shipman also made an order of magnitude estimation (based on the difference in the radiative opacity of $H$ and $He$ gas) and claimed that a $13000K$ DA white dwarf would evolve into a $19000K$ DB white dwarf.  Such estimation is apparently crude since the existence of convection zone is completely ignored.

Later, \citet{1973A&A....27..307B} re-investigated the problem by considering models of convective envelope.  By assuming that the mixing operates at a timescale \emph{much longer} than hydrodynamical timescale and comparing temperature of different envelope models at $q \equiv 1-\frac{M_r}{M_{WD}}=10^{-6}$, they concluded that the $T_{eff}$ of the resultant helium white dwarf cannot exceed $18000K$, i.e., the hottest DB stars cannot be evolved from DA stars.

\citet{1976A&A....52..415K} considered convective mixing in his own white dwarf models and showed \emph{no} increase in $T_{eff}$ upon convective mixing.  He claimed that there should not be an increase in $T_{eff}$ because all envelope solutions converge to the so-called ``radiative zero solution'' independent of the boundary conditions.  Since the publication of \citet{1976A&A....52..415K}, the opinion of the majority seems to be that upon convective mixing, a DA white dwarf will turn into a DC white dwarf at the same $T_{eff}$.

Despite their disagreement in the evolution of $T_{eff}$, all of the above numerical works assume that convective mixing would instantaneously evolve the surface from pure hydrogen composition to nearly pure helium composition.  Such unjustified imposition is probably encouraged by the fact that convection zone in pure helium model is generally, orders of magnitude more massive than that of hydrogen model.  However, as we will show later in this paper, such simplification is not always appropriate.

\citet{1977A&A....61..415V} took a even closer look at the convective mixing picture by investigating the effects of $\mu$-barrier.  Strictly speaking, the white dwarf envelope is not \emph{exactly} layered, since classical diffusion would counteract gravitation-induced diffusion at the $H$/$He$ layer interface \citep{1990ARA&A..28..139D,2004PhR...399....1H}.  Such counteraction would result in a $\mu$-gradient, which stabilizes the convective envelope.  However, It is shown by the authors that $\mu$-barrier cannot stop convective mixing as long as $\frac{m_H}{M_{WD}}>10^{-14}$.

On the observation front, \citet{1990ApJ...351L..21B} analyzed the high Balmer lines of colder DA white dwarfs and found the lines has a tendency to broaden as $T_{eff}$ decreases.  They interpreted this result as an evidence of convective mixing.  Helium gas is spectroscopically invisible at the $T_{eff}$ of their concern ($\lesssim11500K$), hence convective mixing could preserve the original spectral type of a white dwarf (DA) while reducing the surface opacity, resulting in a higher photospheric pressure and broader spectral lines.

Thus, the outcome of convective mixing is rather controversial.  Shall $T_{eff}$ increase after the mixing?  Shall spectral type change?  We will address these questions in the following sections.

\section{Structure of White Dwarf Envelope without Convective mixing} \label{sec:nomix}
Before we investigate the effects of convective mixing, let us briefly study the structure of white dwarf envelope with \emph{fixed} surface composition (i.e., where convective mixing is absent).  A careful study of these relatively simple cases yields insights on the topic we will later address.  We consider two examples: The first being a pure helium white dwarf envelope with $M_{WD}=0.6M_{\sun}$.  The other is a model with identical $M_{WD}$ but has a layer of pure hydrogen ($m_H=10^{-5}M_{\sun}$) on top of the helium layer.  As we described in \S~\ref{sec:intro}, this is the equilibrium configuration of white dwarf envelope in the absence of convective mixing.  

The method we construct our envelope model is standard and similar to that described in \citet{1976ApJS...31..467F}.  However, we mention a few difference:  First, at low $T_{eff}$, we use realistic radiative transfer code of \citet{1999ApJ...520..680H} for surface boundary conditions.  This is a major improvement because envelope structure is very sensitive to the boundary condition at lower $T_{eff}$ \citep{1974ApJ...193..205F}.  Secondly, we implemented the up-to-date EOS (equation of state) table of \citet{1995ApJS...99..713S} and the opacity table of \citet{1992ApJ...401..361R}.  Incidentally, the EOS of Saumon et al. as well as our surface boundary condition subroutine has the ability to deal with $H$--$He$ mixture.  This feature is not exploited in the current section, however, it is of crucial importance as we proceed into section~\ref{sec:atmosphere}.

\begin{figure}
\includegraphics[width=0.5\textwidth]{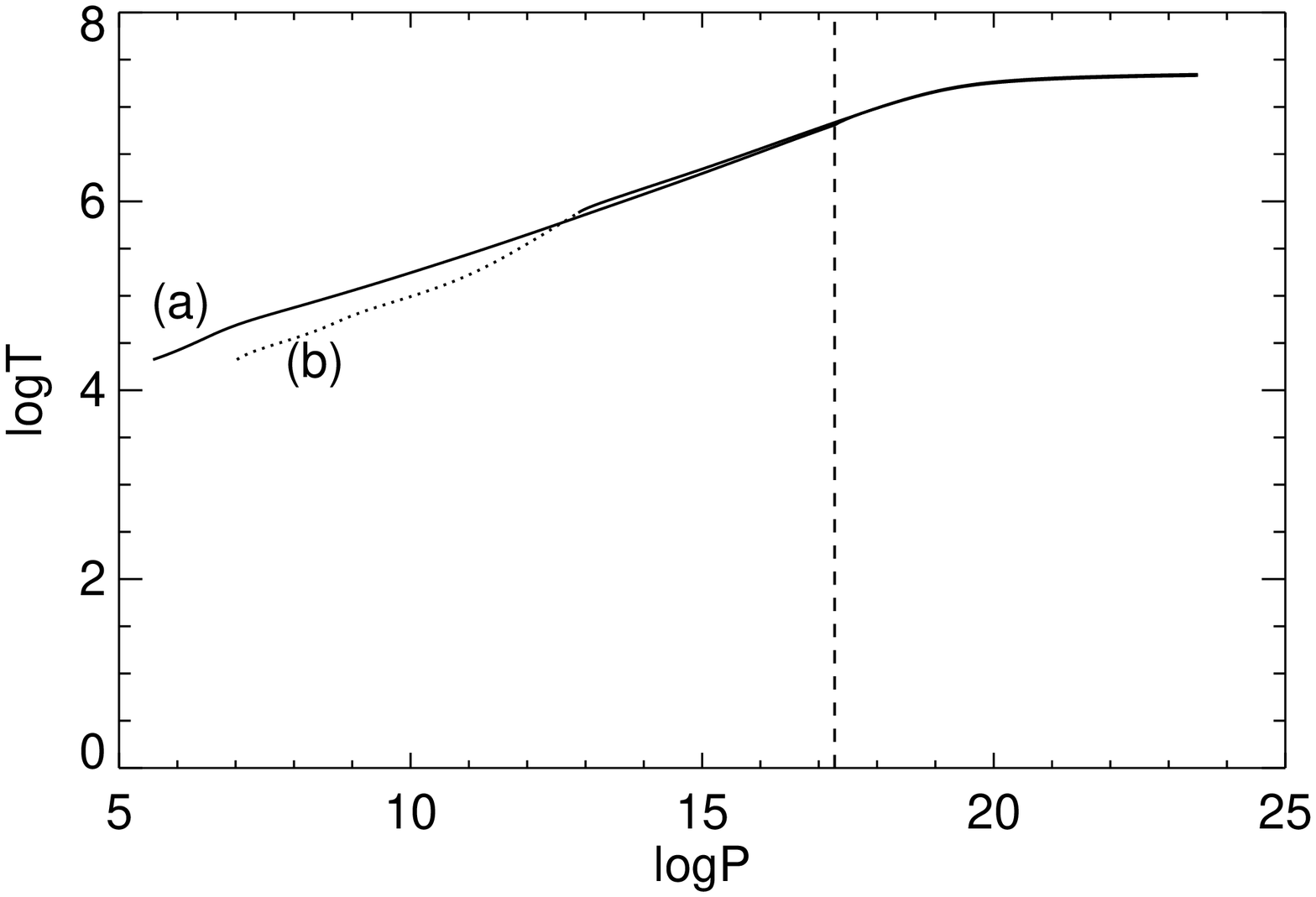}
\includegraphics[width=0.5\textwidth]{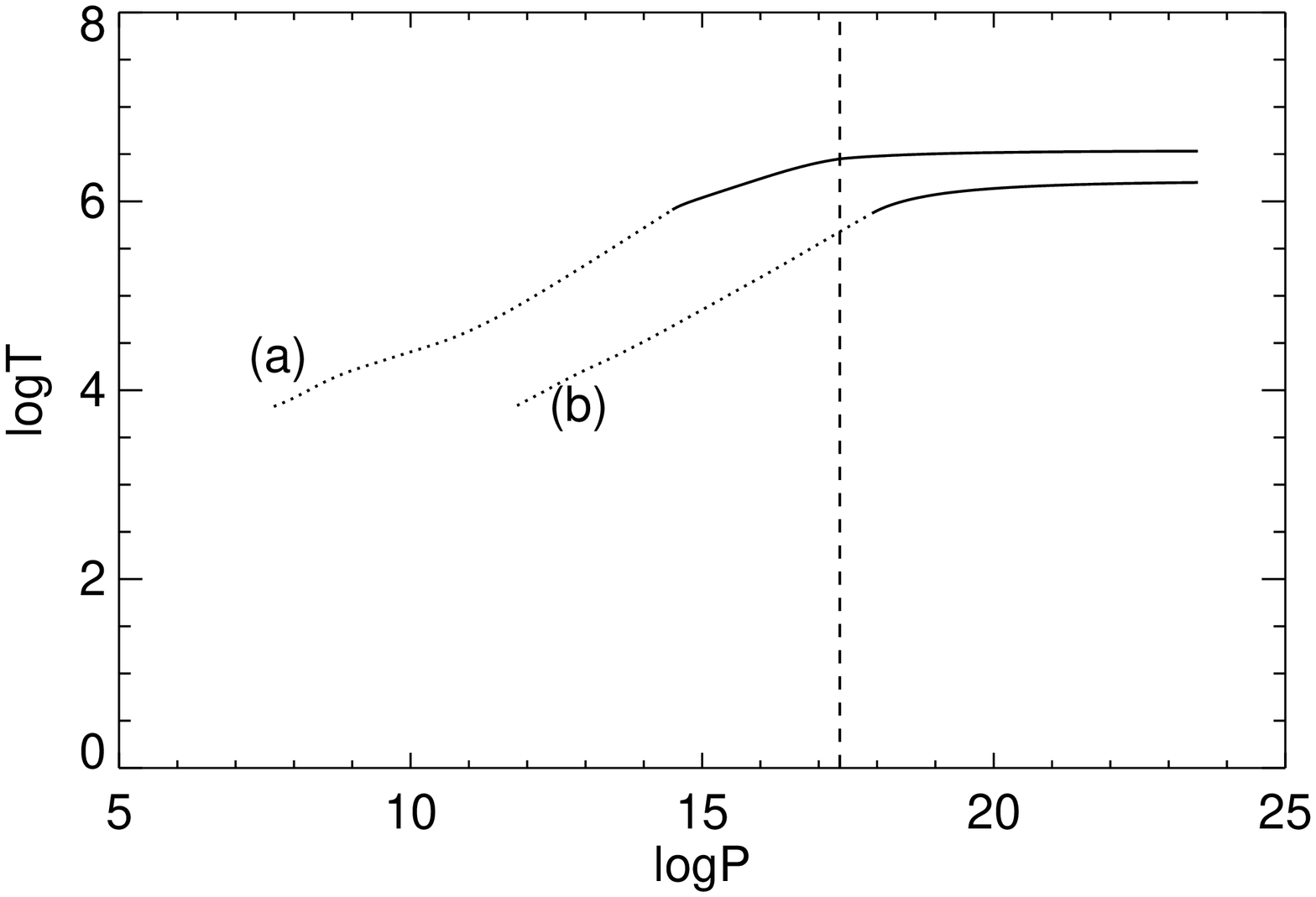}
\caption{Pressure-temperature relation of four exemplary white dwarf envelope models.  Convection zone is represented by dotted lines and radiative/conductive zones are represented by solid lines.  The models with hydrogen surface are marked by ``(a)'' while those with pure helium surface are marked by ``(b)''.  The vertical dashed line indicates the interface between $H$ and $He$ in \emph{hydrogen surface} models and has no meaning pertaining to the helium models.  Upper:  The $P$--$T$ relation at $\log T_{eff}=4.3$.  Lower:  The $P$--$T$ relation at $\log T_{eff}=3.8$.  It is clear that the core temperature $T_c$ is relatively insensitive to the surface boundary condition at high $T_{eff}$, manifesting so-called ``radiative-zero convergence''.}
\label{plotone}
\end{figure}

\begin{figure}
\includegraphics[width=0.5\textwidth]{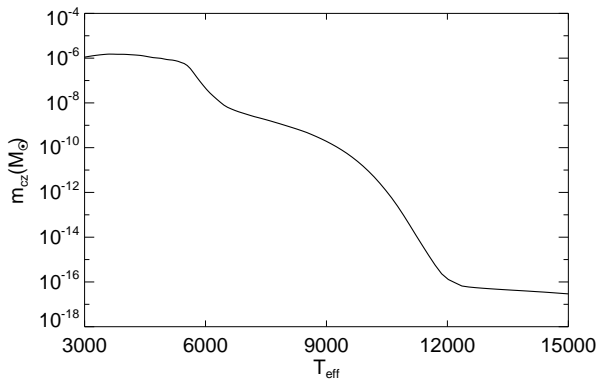}
\caption{$T_{eff}$--$m_{cz}$ relation for a $0.6M_{\sun}$ white dwarf hydrogen envelope.  The convection zone deepens with the decrement of $T_{eff}$ until $T_{eff}\sim4000K$.  The maximum of convection zone mass is around $1.5\times10^{-6}M_\sun$ for a $0.6M_{\sun}$ white dwarf.}
\label{plottwo}
\end{figure}

We calculate the structure of the envelope on a $T_{eff}$ grid ranging from $\log T_{eff}=4.5$ to $\log T_{eff}=3.35$ (all physical quantities in this work are expressed in cgs units unless mentioned otherwise).  A few representative models are shown in figure~\ref{plotone}.  Our results recover the well-known fact that white dwarf envelopes generally feature a surface convection zone on top of radiative/conductive zone which connects to the degenerate core.  The relation between surface convection zone mass ($m_{cz}$) and $T_{eff}$ for hydrogen models are calculated and shown in figure~\ref{plottwo}:  Following the cooling sequence, convection zone first deepens with the decrement of $T_{eff}$.  The curve then flattens and gradually inflected at lower temperatures ($T_{eff}\lesssim4000K$).  Note that the maximum mass of convection zone is $\sim1.5\times10^{-6}M_\sun$, lower than our assumed hydrogen mass $10^{-5}M_\sun$.  Thus, convective mixing needs not to be considered in this example.

\begin{figure}
\includegraphics[width=0.5\textwidth]{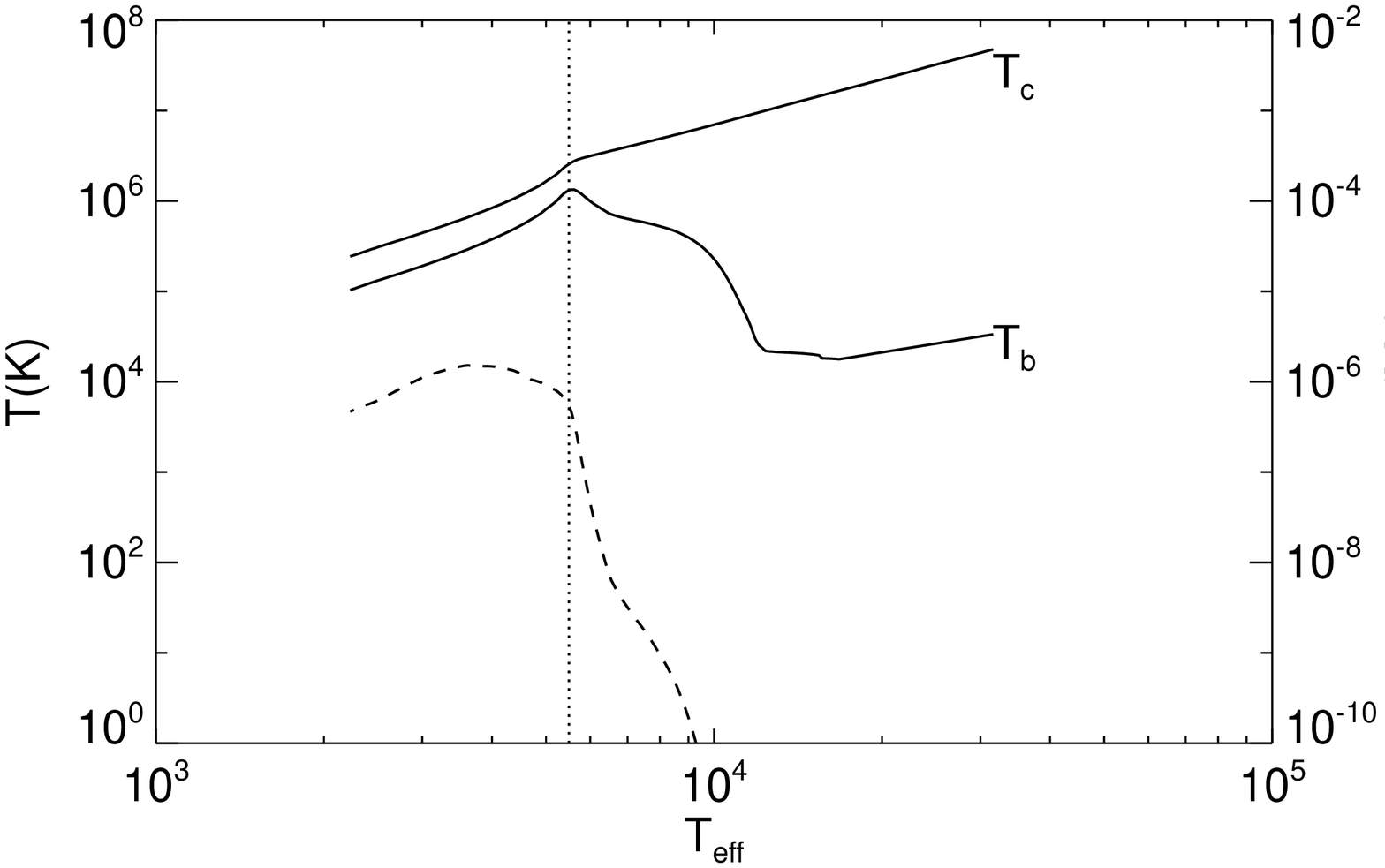}
\caption{$T_{eff}$--$T_c$ and $T_{eff}$--$T_b$ relation for a $0.6M_{\sun}$ hydrogen white dwarf envelope.  Over-plotted on the same figure is $T_{eff}$--$m_{cz}$ relation (dashed line), which we have seen in figure~\ref{plottwo}.  We can see that the $T_{eff}$ where the slope of $T_{eff}$--$T_c$ relation changes coincides with the $T_{eff}$ where the slope of $T_{eff}$--$m_{cz}$ relation flattens (both slope measured in log coordinates).  This $T_{eff}$ is marked by the long vertical dotted line in the figure.}
\label{plotthree}
\end{figure}

We investigate the relation between $T_{eff}$ and core temperature $T_c$ in figure~\ref{plotthree}.  In figure~\ref{plotthree} we see that the relation in log coordinates suddenly changes slope at $T_{eff}\sim4000K$, the same effective temperature where the slope of $T_{eff}$--$m_{cz}$ relation (over-plotted in dashed line for ready reference) changes.  This marks the convective coupling between convection zone and degenerate core, which is documented in \citet{1976ApJS...31..467F}.  Superimposed on the same figure is the relation between $T_{eff}$ and the temperature at the base of convection zone, $T_b$.  It is clear that once the convection zone couples with the degenerate core, the temperature difference between $T_c$ and $T_b$ becomes approximately constant.   

To understand this phenomenon, let us note the opacity of stellar material can be written in the form
\begin{equation}
\kappa=\kappa_0P^nT^{-n-s}
\end{equation}
where $\kappa_0$, $n$ and $s$ are in general \emph{variables}.  However, if they approach constant at the base of convection zone (which is the case when the convection zone couples to the degenerate core--opacity in the core is mainly conductive, c.f., \citet{1974ApJ...193..205F}),  it can be shown that the difference in $\log T$ between the base of convection zone (represented by the subscript $b$) and the stellar core (represented by the subscript $c$) is only a function of the temperature \emph{gradient} at the base of convection zone (which equals to the adiabatic gradient $\nabla_{ad}$ as required by the Schwarzschild criterion) as follows:
\begin{eqnarray}
\Delta \log T = \int_{\log P_b}^{\log P_c}\frac{d\log T}{d\log P} d \log P \\
\approx \int_{\log P_b}^\infty \frac{d\log T}{d\log P} d \log P = \int_{\log P_b}^\infty \frac{3L}{16\pi acGM} \frac{\kappa P}{T^4} d \log P \\
= \left.\nabla_{ad}\right|_{\log P=\log P_b} \int_{\log P_b}^\infty (\frac{P}{P_b})^{n+1}(\frac{T}{T_b})^{-(n+s+4)} d \log P \\
= \left.\nabla_{ad}\right|_{\log P=\log P_b} \int_{\log P_b}^\infty 10^{(n+1)(\log P-\log P_b)-(n+s+4)(\log T-\log T_b)} d \log P
\end{eqnarray}
In the second line we have changed the upper limit of integration from $\log P_c$ to $\infty$ because the integrand is already vanishing small at $P \sim P_c$ (manifested by the fact that $\log T-\log P$ relation is horizontal as shown in figure~\ref{plotone}).  The whole integration is done in the conductive/radiative region so the formula of radiative temperature gradient is used.  We represent the radiative/conductive temperature gradient (i.e., $\frac{d\log T}{d\log P}$) by $\nabla$ and Taylor expand $\log T$ at the base of convection zone:
\begin{eqnarray}
\log T=\log T_b + (\log P-\log P_b)\left.\nabla\right|_{\log P=\log P_b}  + (\log P-\log P_b)^2\:\left.\frac{1}{2}\frac{d\,\nabla}{d\log P}\right|_{\log P=\log P_b} +\dots
\end{eqnarray}
We also note that all derivatives of $\nabla$ with respect to pressure (as well as that of any functions of $\nabla$) at the base of convection zone are functions of $\left. \nabla_{ad} \right|_{\log P=\log P_b}$. (For compactness, we hereafter denote the adiabatic gradient evaluated at the base of convection zone simply by $\nabla_{ad}$, dropping the evaluation notation whenever the context is clear):
\begin{eqnarray}
\left.\frac{d\,\nabla}{d\log P}\right|_{\log P=\log P_b}=\left( \left.\frac{\partial\,\nabla}{\partial \log P}\right|_{\log T} +
\left.\frac{\partial\,\nabla}{\partial \log T}\right|_{\log P}\nabla \right)_{\log P=\log P_b}\\
=\left[(n+1)\nabla-(n+s+4)\nabla^2\right]_{\log P=\log P_b}=(n+1)\nabla_{ad}-(n+s+4)\nabla^2_{ad} \\
\left.\frac{d\,f(\nabla)}{d\log P}\right|_{\log P=\log P_b}=\left(f'(\nabla)\frac{d\,\nabla}{d\log P}\right)_{\log P=\log P_b}=f'(\nabla_{ad})((n+1)\nabla_{ad}-(n+s+4)\nabla^2_{ad})
\end{eqnarray}

\begin{figure}
\includegraphics[width=0.5\textwidth]{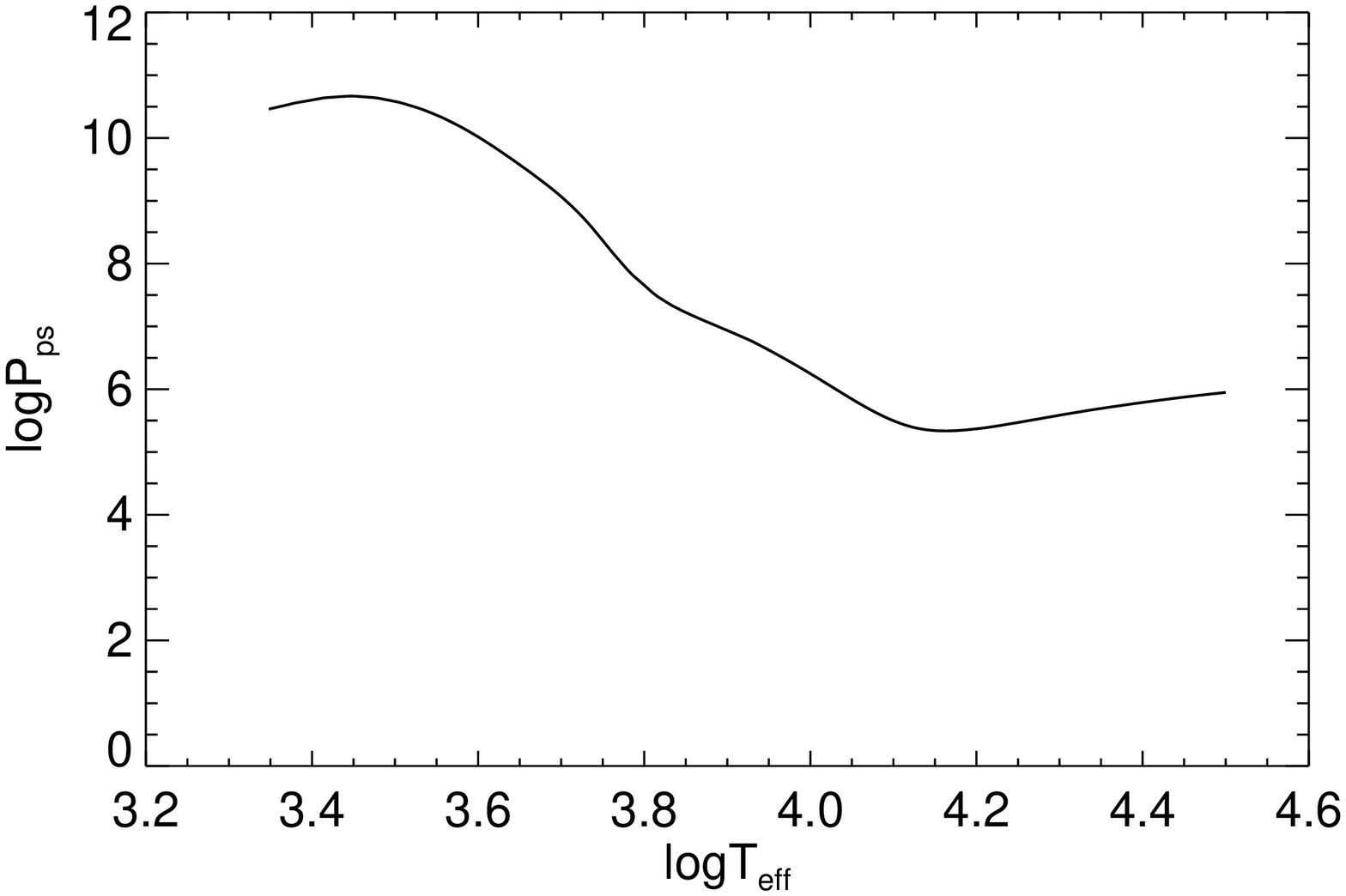}
\includegraphics[width=0.5\textwidth]{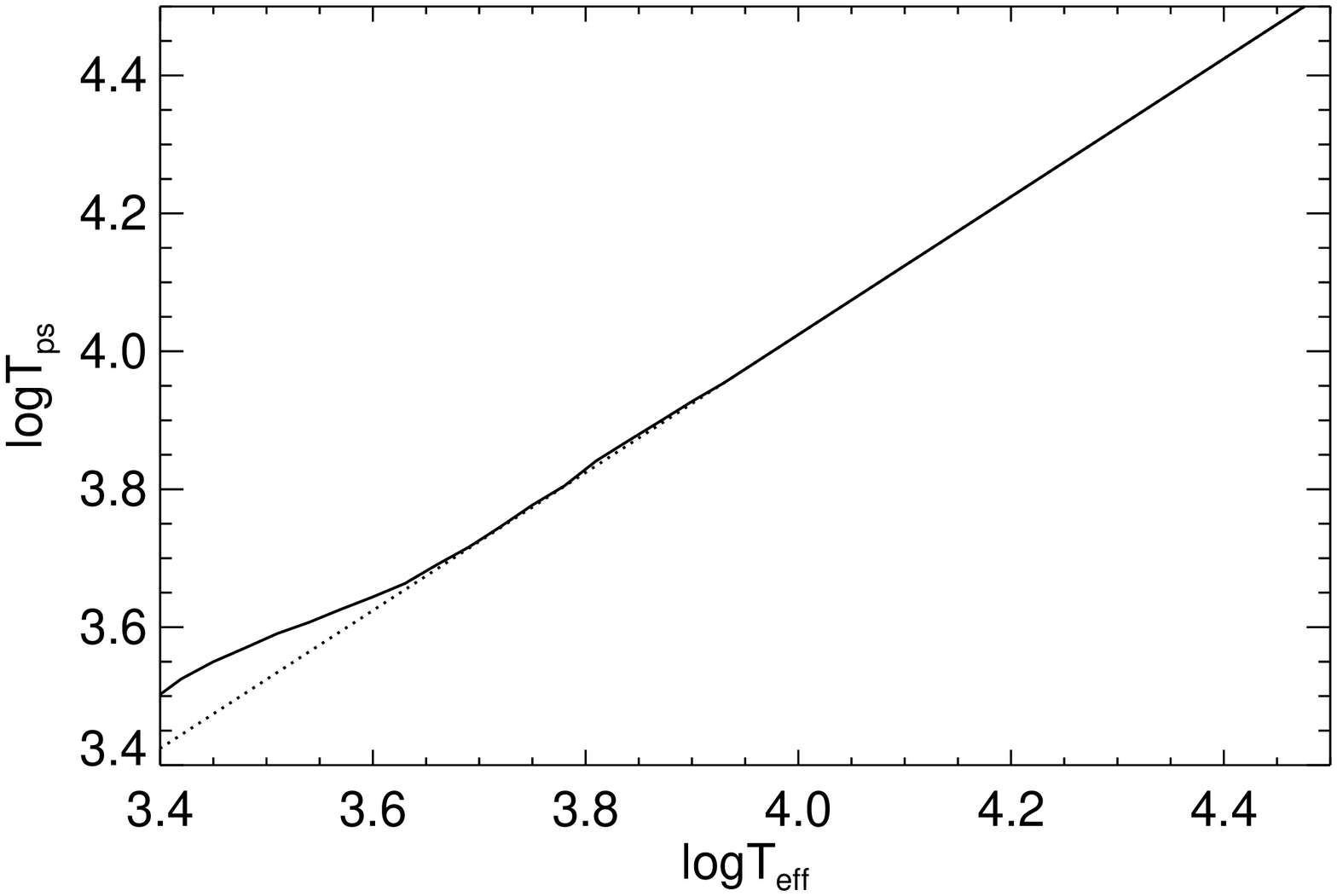}
\caption{The $T_{eff}$--$P_{ps}$ relation and $T_{eff}$--$T_{ps}$ relation of pure hydrogen photosphere obtained from the radiative transfer code of \citet{1999ApJ...520..680H}.  Upper: $T_{eff}-P_{ps}$ relation.  Lower: $T_{eff}-T_{ps}$ relation.  The dotted line is the Eddington $T$--$\tau$ relation $T^4=\frac{3}{4}T_{eff}^4(\tau+\frac{2}{3})$ taken at $\tau=1$.}
\label{plotfour}
\end{figure}

Switching the integration variable to
\begin{equation}
\xi \equiv \log P-\log P_b
\end{equation}
It follows that
\begin{eqnarray}
\Delta \log T \approx \int_0^\infty\nabla_{ad}\;10^{g(\xi,\nabla_{ad})}d\xi \equiv G(\nabla_{ad})
\end{eqnarray}
This explains the fact that $\Delta \log T$ is only a function of $\nabla_{ad}$.  Since the $\nabla_{ad}$ at the base of convection zone is approximately constant along the cooling sequence, so is $\Delta \log T$.  Thus, in the convective coupling regime, we are able to relate $T_{eff}$ and $T_c$ through the following formula:
\begin{eqnarray}\label{eq:rejuvenation}
\log T_c-\log T_{ps}\approx\int_{\log P_{ps}}^{\log P_b} \nabla_{ad} \;d\log P + \left.G(\nabla_{ad})\right|_{\log P=\log P_b}\\ =(\log P_b - \log P_{ps})\langle\nabla_{ad}\rangle+G(\nabla_{ad})
\end{eqnarray}
Where the subscript $ps$ stands for the (base of) photosphere.  $T_{ps}$ and $P_{ps}$ as functions of $T_{eff}$ are found numerically through the radiative transfer code of \citet{1999ApJ...520..680H} and is shown in figure~\ref{plotfour}.  All of our models show high efficiency of convective transport in the envelope and one need not to distinguish between the convective temperature gradient with $\nabla_{ad}$.  
 
\begin{figure}
\includegraphics[width=0.5\textwidth]{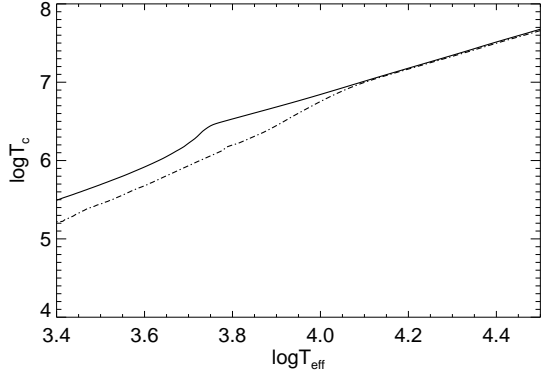}
\caption{$T_{eff}$--$T_c$ relation for DA(solid) and DB(dash-dotted) white dwarfs.}
\label{plotfive}
\end{figure}

Figure~\ref{plotfive} plots $T_{eff}$--$T_c$ relation for both hydrogen and helium model on the same figure.  We see that at $T_{eff}\gtrsim12000K$, the two relations are almost identical, demonstrating the ``radiative-zero convergence'' (c.f. \citet{1976A&A....52..415K,1976ApJS...31..467F}) when the base of convection zone and the conductive core is buffered by a radiative zone for both models.  In fact, we have already seen an indication of this phenomenon in the upper panel of figure~\ref{plotone}.  The $T_{eff}$--$T_c$ relation is thus divided into two different regimes:  A high $T_{eff}$ regime where $T_{c}$ is insensitive to the atmospheric chemical composition, and a low $T_{eff}$ regime where equation~\ref{eq:rejuvenation} starts to hold for models with core-coupled convection zone.

We are now equipped with the knowledge of atmosphere of two extreme surface composition.  In the following sections, we will address the complications due to convective mixing.

\section{Self-consistent Treatment of Convective Mixing Envelopes}\label{sec:atmosphere}
As we have mentioned in \S~\ref{sec:Mix}, most calculations in the current literature assumes the hydrogen in post-mixing white dwarf envelope to be highly diluted and has negligible influence on envelope structure.  As such, both the opacity and equation of state are taken to be that of helium.  This assumption is not self-consistent in the sense that it forces the total hydrogen mass ($m_H$) to go from a finite value to zero after convective mixing.  Considering that the opacity of hydrogen is much larger than that of helium around the $T_{eff}$ of interest, this assumption needs to be tested.  Here we take a more rigorous approach to ensure self-consistency in hydrogen content.

Instead of assuming the chemical composition to go from pure hydrogen to pure helium, we assign the post-mixing white dwarf envelopes to possess a (yet \emph{unknown}) surface composition $X=X_{surf}$, where $X$ is the hydrogen fraction.  

We take the convective mixing white dwarf atmosphere to be a two layer structure, i.e., a homogeneous layer of \emph{convective} $H$--$He$ mixture (with $X=X_{surf}$) on top of a helium layer, which is a direct generalization of the DA white dwarf structure under the influence of convective instability (c.f. \S~1).  The rationale lies in the fact that convective mixing is by far the strongest counter-separation mechanism, and diffusion cannot possibly separate the stellar material within its range of influence.  Besides, we only consider white dwarfs with $10^{-11}M_\sun<m_H<10^{-7}M_\sun$ to avoid the possible complications caused by classical diffusion (such as prevention and/or delay in convective mixing, c.f. \citep{1977A&A....61..415V}).  Thus, the chemical structure of a convective mixing white dwarf is described by a step function:
\begin{eqnarray}\label{chemicalprofile}
X=X_{surf}, \ \ \ {\rm if\ } P<\frac{m_H}{X_{surf}}\frac{g}{4\pi R^2} \\
X=0, \ \ \ {\rm if\ } P>\frac{m_H}{X_{surf}}\frac{g}{4\pi R^2}
\end{eqnarray}
where the location of the step (i.e., chemical composition inhomogeneity interface) ought to be matched with the boundary of convection zone.

\begin{figure}
\includegraphics[width=0.5\textwidth]{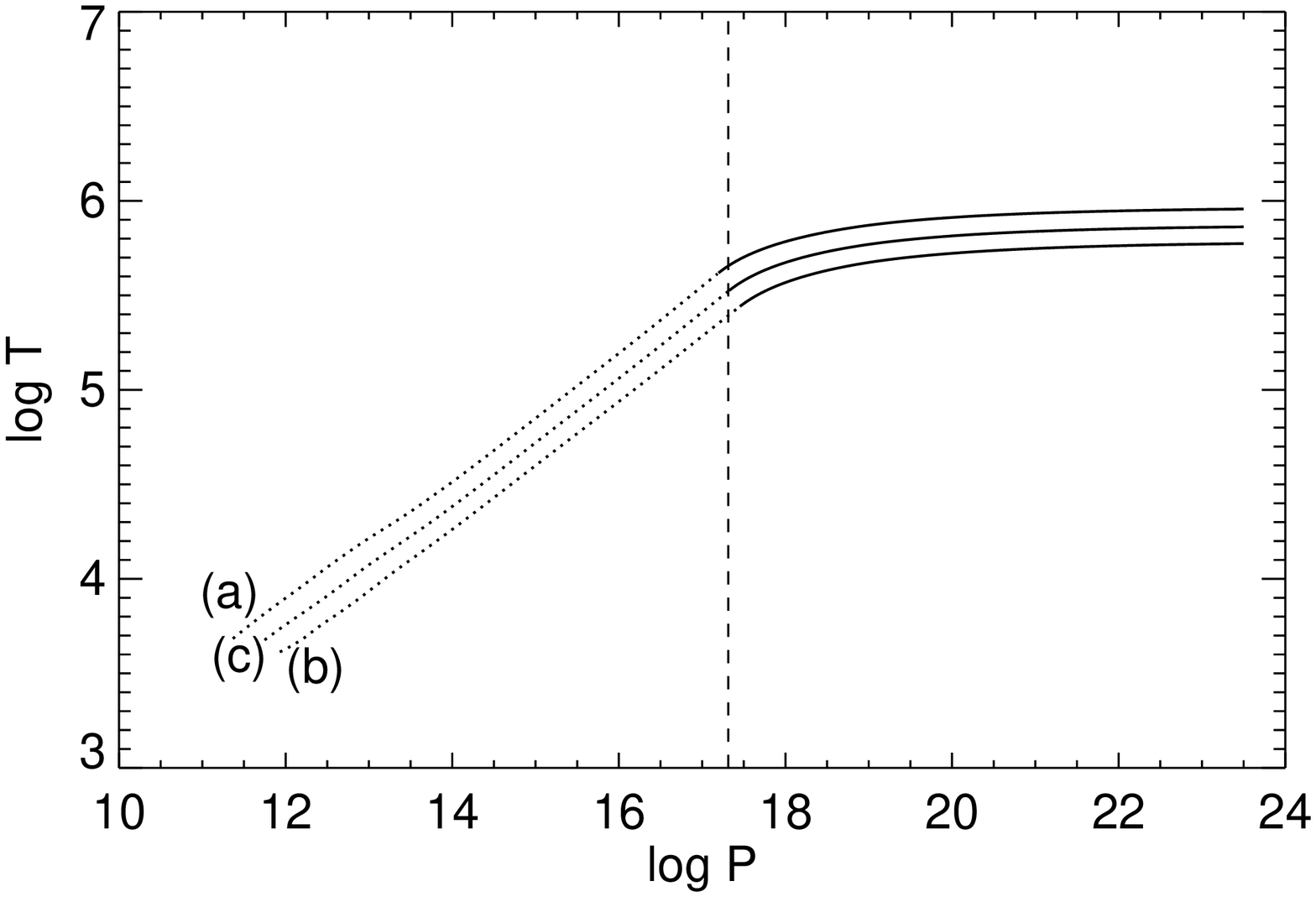}
\caption{Three white dwarf envelope models with $X_{surf}=0.005$ and $m_H=4.0\times10^{-8}M_\sun$.  The chemical composition interface is determined by $X_{surf}$ and $m_H$ and is denoted by the long vertical dashed line.  The three models have different $T_{eff}$.  In (a), $T_{eff}=4500K$.  The resultant convection zone does not reach the interface, yielding an inconsistency.  In (b), $T_{eff}=3800K$.  The resultant convection zone penetrates the interface, yielding another inconsistency.  However, in (c), $T_{eff}=4128K$, the boundary of surface convection zone and the chemical composition interface match.  The conclusion we draw here is the following:  If there is a white dwarf with $X_{surf}=0.005$ and $m_H=4.0\times10^{-8}M_\sun$, then $T_{eff}$ can only be $4128K$ at this neighborhood.}
\label{plotseven}
\end{figure}

To proceed, we calculate the envelope structure on a \emph{grid} of $T_{eff}$ and $X_{surf}$ for a given $m_H$ with the routine described in \S~\ref{sec:nomix}, which features detailed treatment on the boundary condition and EOS of $H$--$He$ \emph{mixture}.  The resultant models (figure~\ref{plotseven}) have structure similar to that of non-mixing envelopes (i.e., possess surface convection zone on top of radiative/conductive zone), however, the boundary of surface convection zone generally mismatch with the chemical composition interface (which is designated by equation~\ref{chemicalprofile} prior to the envelope integration).  Our task of searching candidates of convective mixing envelopes is thus tantamount to picking envelope parameters ($m_H, M_{WD}, X_{surf}, T_{eff}$) which allows a matching between the interface of chemical composition and the boundary of surface convection zone.

A set of concrete examples is provided in figure~\ref{plotseven}.  In line (a) of figure \ref{plotseven} we solved the structure of a $0.6M_{\sun}$ white dwarf with $m_H=4.0\times10^{-8}M_{\sun}$, $X_{surf}=0.005$ at $T_{eff}=4500K$.  The line in the plot depicts the pressure-temperature relation in that white dwarf envelope.  The dotted line denotes the convection zone while the solid line denotes the radiative zone.  According to our prescribed $X_{surf}$ and $m_H$, the interface between surface layer and underlying helium layer locates at $\log P=\log \frac{m_H}{X_{surf}}\frac{g}{4\pi R^2}=17.3$, which is indicated by the long vertical dashed line.  We found that this model cannot be a candidate of convective mixing envelope because while the layer of $X_{surf}=0.005$ extends to $\log P=17.3$, the convection zone merely reaches $\log P=17.18$.  This configuration cannot be realized because hydrogen in such envelope would diffuse upwards, altering the prescribed chemical profile.  In other words, the convection zone mass in the resultant model underestimates the self-consistent convection zone and we do not consider this model as a candidate of self-consistent, convective mixing envelope.

In line (b) of figure \ref{plotseven} we solved white dwarf atmosphere with identical $m_H$ and $X_{surf}$, but $T_{eff}=3800K$.  The resultant model cannot be realized as well because the convection zone now extends to $\log P=17.44$, which is below the $H$--$He$ mixture/$He$ layer interface.  The convective motion would mix pure helium gas with the gas of $X=0.005$, resulting in a gas which is more helium-rich.  In this case, the convection zone mass in our model overestimates that of self-consistent models.  However, a self-consistent model can be found between the two mentioned cases with $T_{eff}\sim4128K$ (line (c)).

In mathematical terms, the parameter that allows self-consistent convective mixing envelopes is defined by the contour:
\begin{equation}
m_{cz}(T_{eff},X_{surf};M_{WD},m_H)=\frac{m_H}{X_{surf}}
\end{equation}

\begin{figure}
\includegraphics[width=0.45\textwidth]{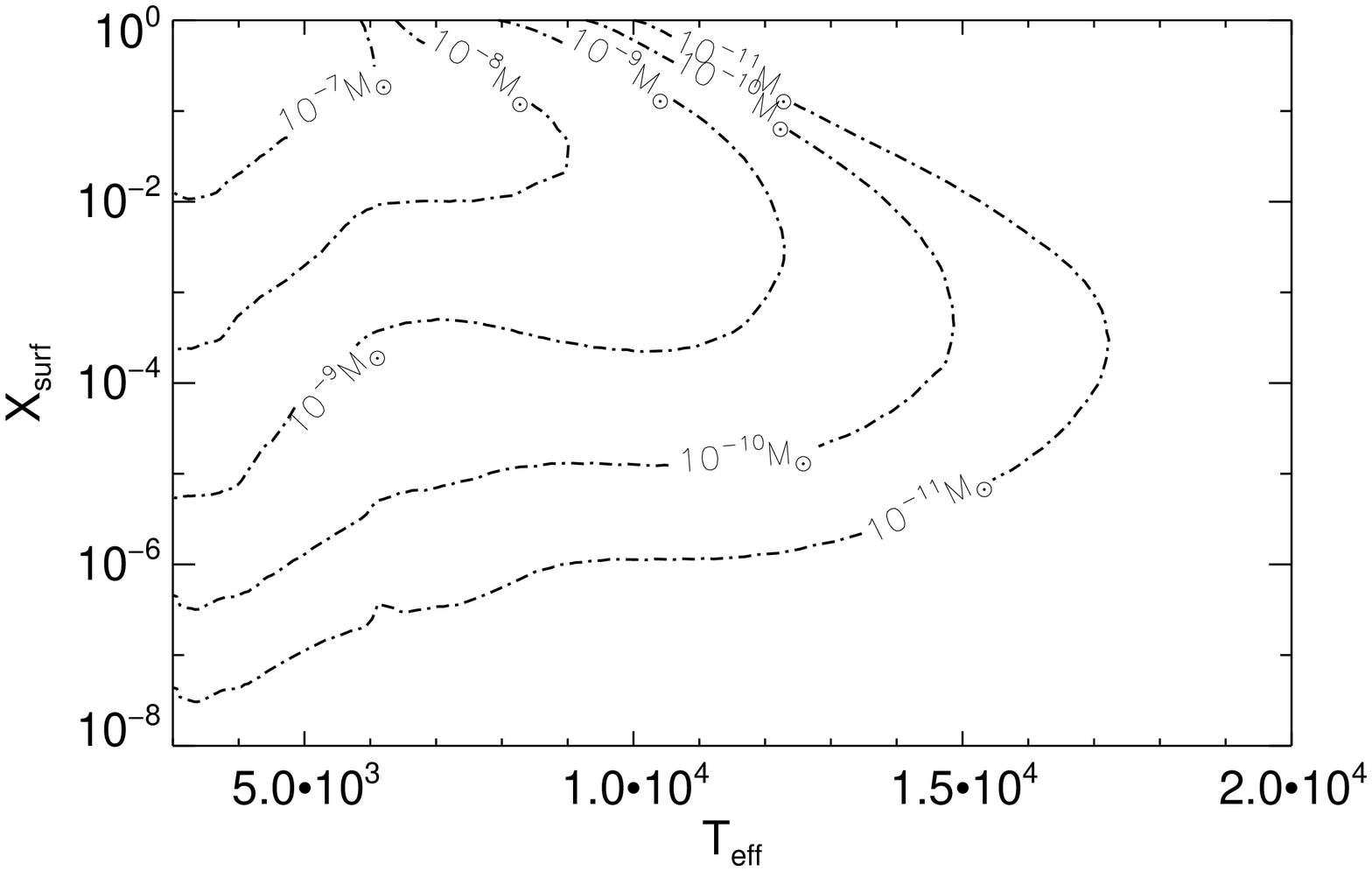}
\includegraphics[width=0.45\textwidth]{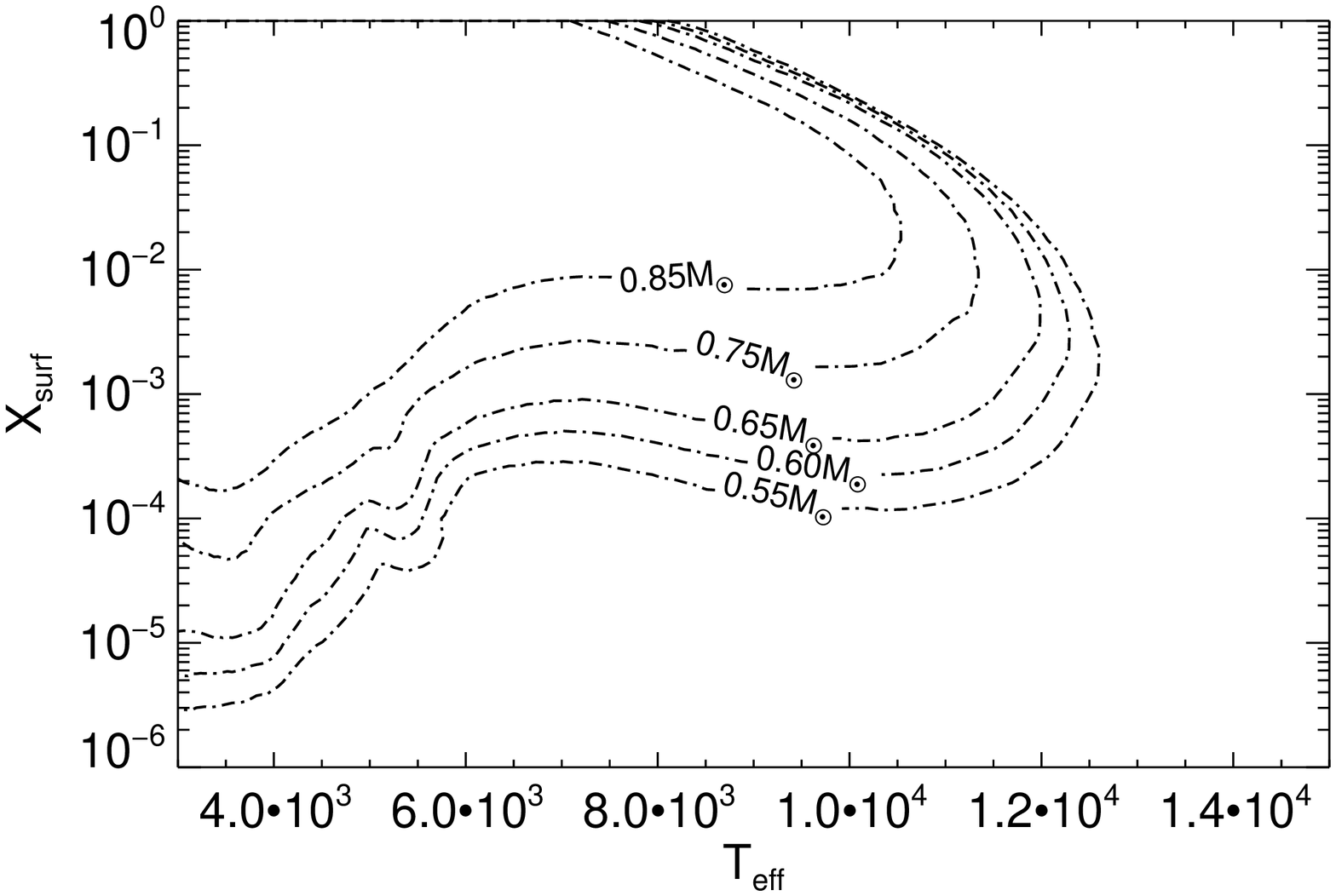}
\caption{Upper: The contour of hydrogen conservation for a $0.6M_\sun$ white dwarf with various $m_H$ (value marked on the contour line).  Lower: The contour of hydrogen conservation for white dwarf of various mass (value marked on the contour line) with $m_H=10^{-9}M_\sun$.}
\label{ploteight}
\end{figure}

We have calculated the contour for a wide range of $m_H$ and $M_{WD}$.  Some representative contours are plotted in figure~\ref{ploteight} (For the rest of this paper, such contour is referred as a ``contour of hydrogen conservation'').  We clearly see that the possibility for convective mixing envelope to possess a non-negligible surface hydrogen fraction.  Therefore, the assumption of highly diluted surface composition in convective mixing envelopes is not always proper.

To conclude, we have placed a constraint on the possible outcomes of convective mixing:  Namely, the post-mixing envelope must lie on the contour of hydrogen conservation in order to conserve the total hydrogen mass.  In the following sections, we will exploit this result and derive the evolutionary sequence, cooling curves and chemical evolutionary curves for convective mixing white dwarfs.

\section{Revisiting the Convective Mixing Scenario}
\label{sec:sequence}
We now formally revisit the convective mixing scenario in white dwarfs with intermediate amount of hydrogen.  As we have reviewed in \S~\ref{sec:Mix}, the surface convection zone in a hot hydrogen white dwarf deepens continuously as $T_{eff}$ decreases.  Its spectrum remains DA until the lower boundary of convection zone reaches the $H$/$He$ boundary.  At this point, convective mixing occurs and the $X_{surf}$ is bound to change. We claim the outcome is governed by two rules:
\begin{enumerate}
\item The total amount of hydrogen must be conserved.  This rule can be met by requiring the post-mixing envelope to lie on the contour of hydrogen conservation.
\item\label{item:Tcprinciple} $T_c$ is approximately invariant.  The underlying principle of this rule is \emph{energy conservation}.  White dwarf is a highly degenerate system, therefore, the gravitational potential energy released during convective mixing is small compared to the total thermal energy, which is represented by $T_c$.
\end{enumerate}
It also follows from the reasoning of principle~\ref{item:Tcprinciple} that $T_c$ is monotonically decreasing over time (because heat energy is continuously released through radiation).  It can therefore serve as a sorting index and we proceed to calculate the $T_c$ for all models that are possible to realize during the evolutionary sequence.  This include all pure DA envelopes with $T_{eff}>T_{mix}$ (the convective mixing $T_{eff}$) and the convective mixing envelopes that lie on the contour of hydrogen conservation.

\begin{figure}
\includegraphics[width=0.5\textwidth]{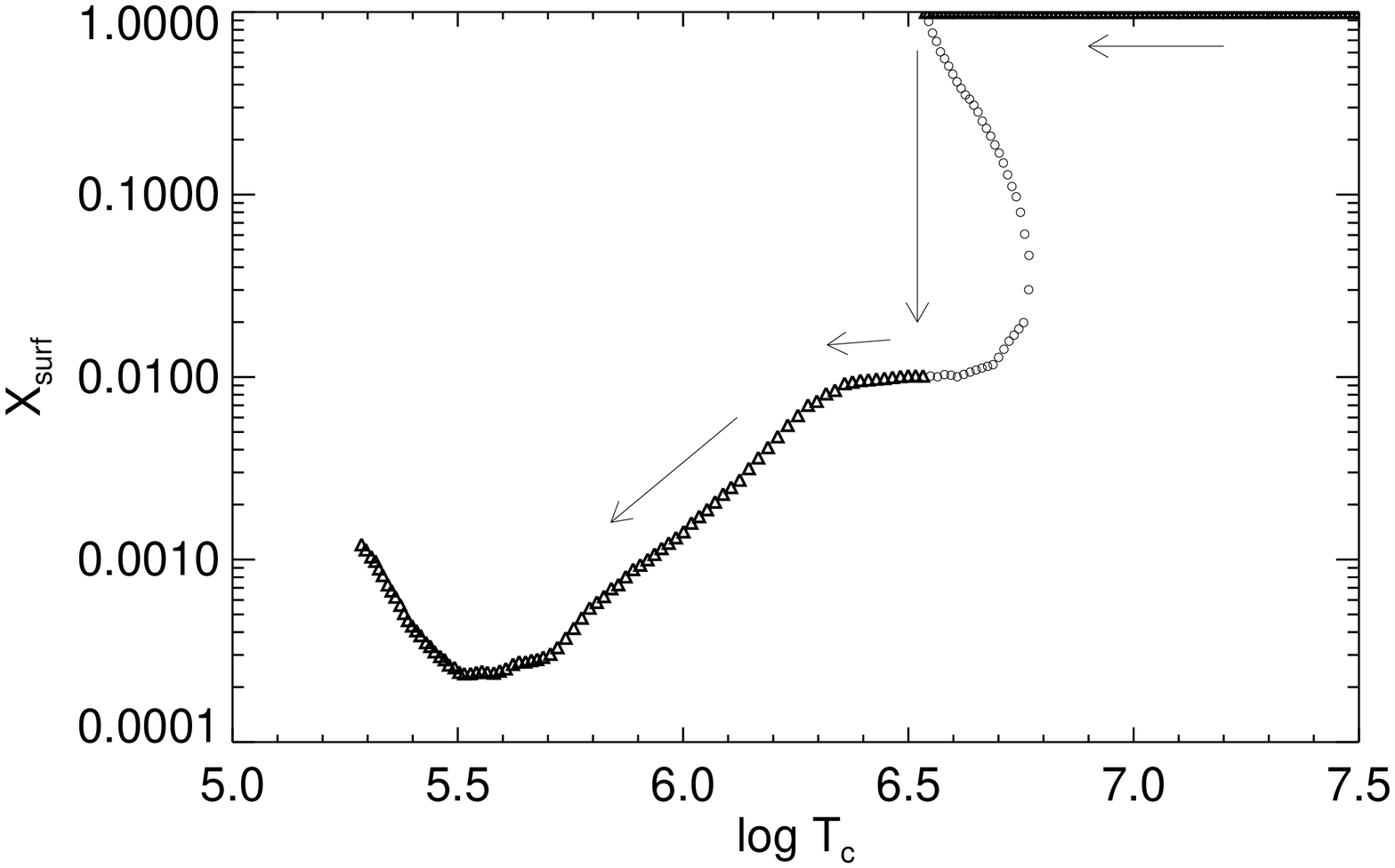}
\includegraphics[width=0.5\textwidth]{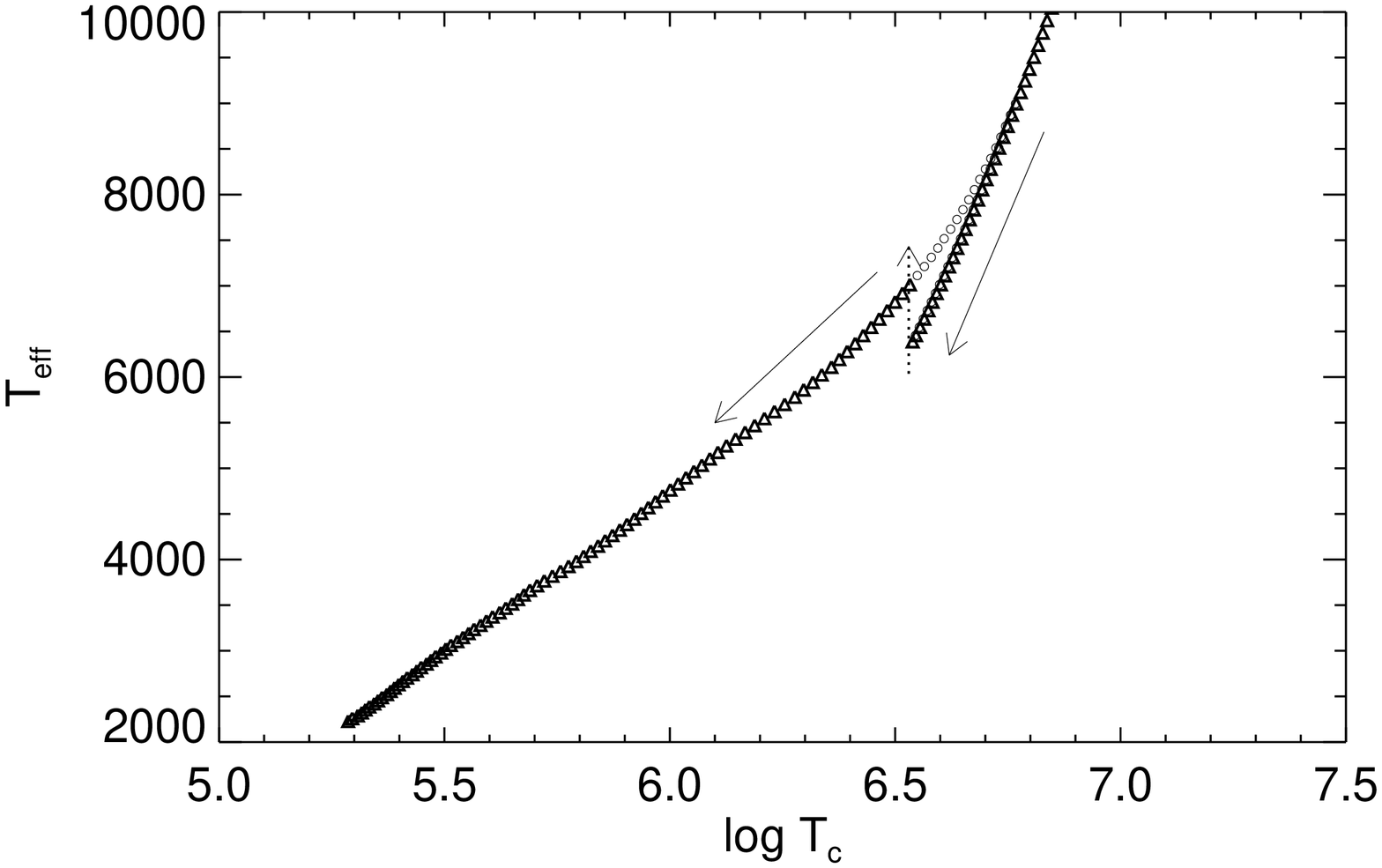}
\includegraphics[width=0.5\textwidth]{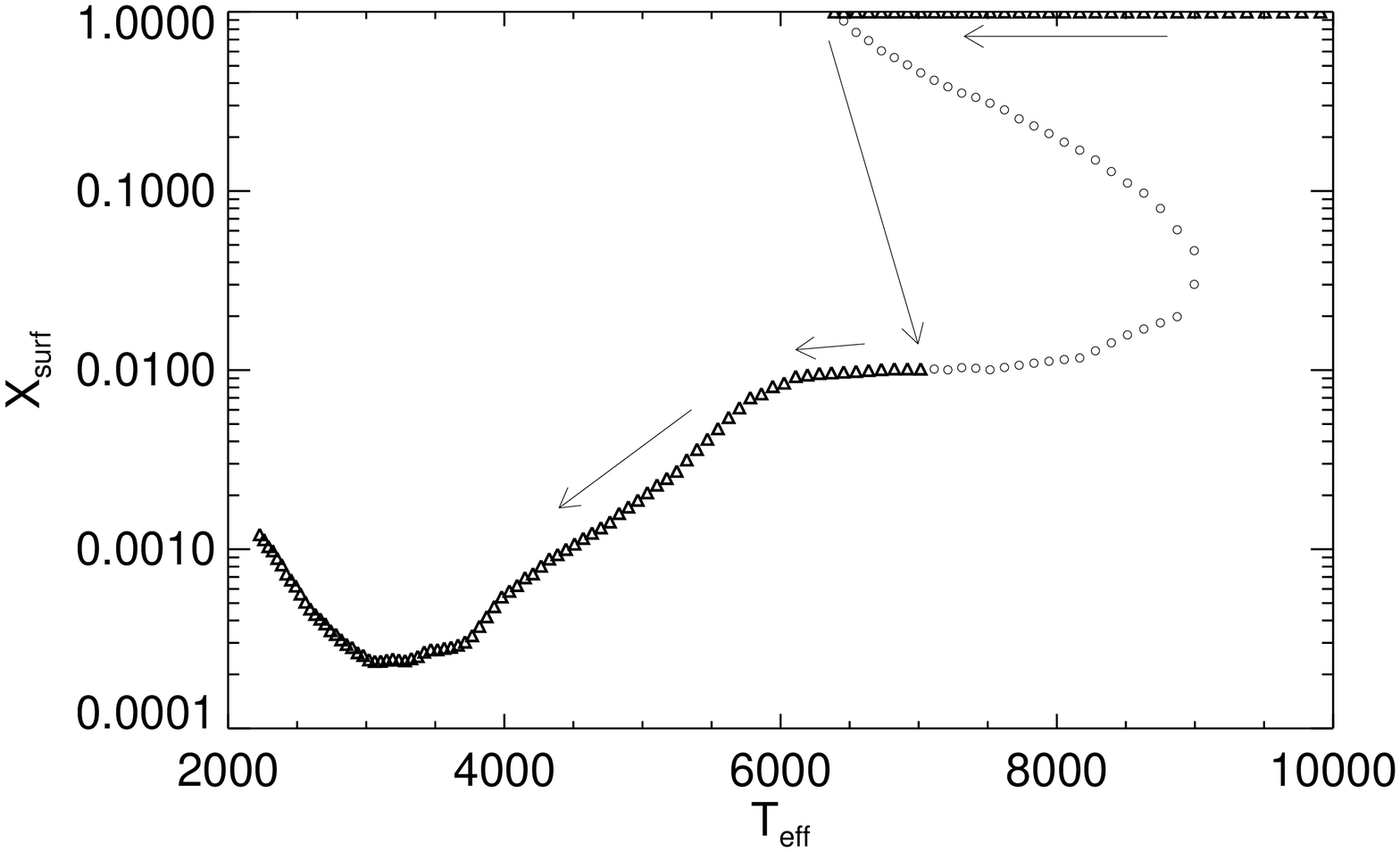}
\caption{$T_c$--$T_{eff}$ relation (upper panel), $T_c$--$X_{surf}$ relation (middle panel) and $T_{eff}$--$T_c$ relation (lower panel) for white dwarf envelopes with a total hydrogen amount of $10^{-8}M_\sun$ from self-consistent, convective mixing calculation.  The models which will be realized during the cooling sequence are marked by triangular symbols while those will not are marked by hollow circles.  The arrows indicate the track of evolution, based on $T_c$.  We can see in the lower panel that the track is zig-zag in $T_{eff}$--$X_{surf}$ space.}
\label{plotnine}
\end{figure}

An exemplary set of results is summarized in figure~\ref{plotnine}, where the relations between $T_c$, $T_{eff}$ and $X_{surf}$ is plotted for $\sim200$ candidate models.

It is important to note, however, that the contour of hydrogen conservation is merely a constraint on possible outcomes.  Models on the contours need \emph{not} to be realized in the cooling sequence.  Let us denote the core temperature of the coolest pure DA star (among the candidate models) to be $T_c^\ast$.  This particular model, as we have reviewed, is just about to be convectively-mixed.  We also know that prior to this model, convective mixing is not present and the white dwarf has pure hydrogen surface, due to the efficient separation of diffusion.  Therefore, all models on the contour of hydrogen conservation (i.e., those does \emph{not} have pure hydrogen surface) with $T_c>T_c^\ast$ should be excluded from the cooling sequence.

The evolution of white dwarf in $T_{eff}$ and $X_{surf}$ is thus found by sorting the models according to $T_c$ after the exclusion of unrealized models.  It is indicated by the arrows in upper and middle panels of figure \ref{plotnine}, where we have marked the excluded models in different symbol.  In our example ($m_H=10^{-8}M_\sun$), it is apparent that $X_{surf}$ decreases and $T_{eff}$ increases upon convective mixing.  The evolution track of white dwarf in $T_{eff}$-$X_{surf}$ space is therefore zigzag, as shown in the lower panel of figure \ref{plotnine}.  The white dwarf travels through a segment of $T_{eff}$ interval twice during its course of evolution.  The first time as a pure DA star, the second time as a white dwarf with lower $X_{surf}$.  Incidentally, we have examined the behavior of $m_H(T_{eff},X_{surf})$ to ensure the stability of our solutions.

\section{The Cooling curves and the Chemical Evolution curves}
\label{sec:evolve}
In order to obtain the cooling curve ($t$--$T_{eff}$ relation) and the chemical evolution curve ($t$--$X_{surf}$ relation), one needs not only to know the evolution track in the parameter space (that constructed by $T_{eff}$, $T_c$ and $X_{surf}$) but also needs to know the time spent between different evolutionary stages.  This sets the goal of this section.

\begin{figure}
\includegraphics[width=0.5\textwidth]{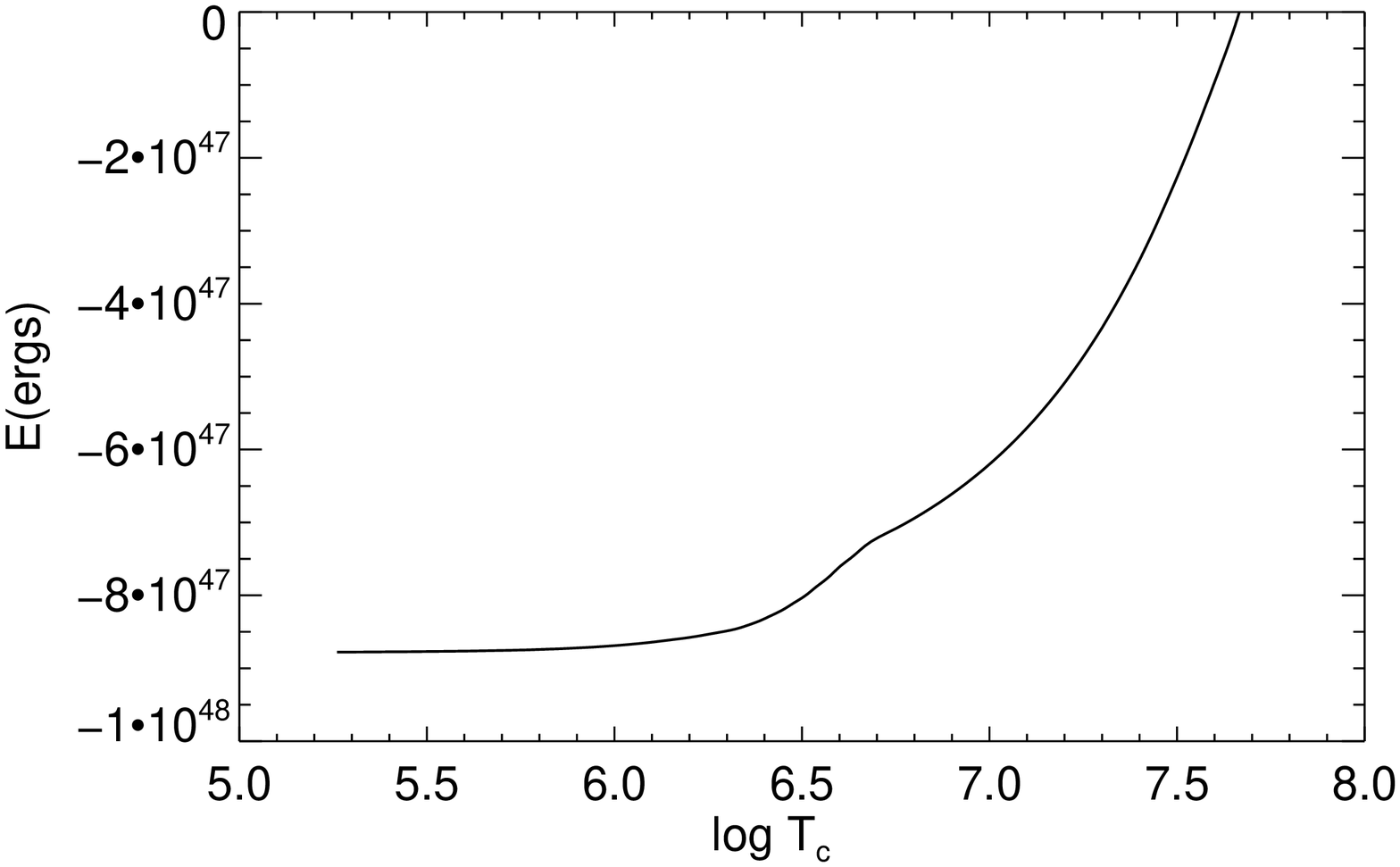}
\caption{The relation between $\log T_c$ and thermal energy.  The zero point of energy is set at the beginning of the evolution sequence, where $\log T_c=7.7$ (hence the negative value of energy).}
\label{plotten}
\end{figure}

With the help of a full evolutionary model of white dwarfs \citep{1999ApJ...520..680H}, we are able to calculate the energy of a white dwarf as a function of its core temperature, i.e., $E=E(T_c)$.  The evolutionary model of \citet{1999ApJ...520..680H} is designed to deal with models of a fixed atmospheric composition.  However, it is sufficient for our purposes since the energy content in the atmosphere is negligibly small.  Our result is shown in figure \ref{plotten}.

The cooling time of a particular evolutionary sequence can be calculated by relating $T_{eff}$ to the core temperature and energy content:

\begin{eqnarray}
dt=-\frac{dE(T_c)}{4\pi R^2 \sigma T^4_{eff}(T_c)} \\\label{tEL}
t(T_c)=\int^{E(T_c)}  -\frac{dE}{4\pi R^2 \sigma T^4_{eff}}= \int^{T_c}  \frac{-\frac{dE(T_c)}{dT_c}dT_c}{4\pi R^2 \sigma T^4_{eff}}\label{ttc}
\end{eqnarray}

Note that $E(T_c)$ is scenario independent whereas $T_{eff}(T_c)$ depends on $X_{surf}$ and is specific to the particular evolutionary sequence.

We are now able to calculate the chemical evolution curve and $T_{eff}$ evolutionary curve in the following way:
\begin{enumerate}
\item For a given $m_H$, we first identify the evolutionary sequence of a white dwarf in the convective mixing scenario.  $T_{eff}(T_c)$ and $X_{surf}(T_c)$ is thus obtained.  This process is described in section~\ref{sec:sequence} and an exemplary result was shown in figure \ref{plotnine}.
\item After the evolutionary sequence is identified, we use $E(T_c)$ from the full evolutionary model of \citet{1999ApJ...520..680H} and $T_{eff}(T_c)$ to calculate the cooling time $t(T_c)$ through equation~\ref{tEL}.
\item $T_{eff}(t)$ and $X_{surf}(t)$ are then obtained by inverting $t(T_c)$ and substituting it into $T_{eff}(T_c)$ and $X_{surf}(T_c)$.
\end{enumerate}

\begin{figure}
\includegraphics[width=0.5\textwidth,height=0.3\textheight]{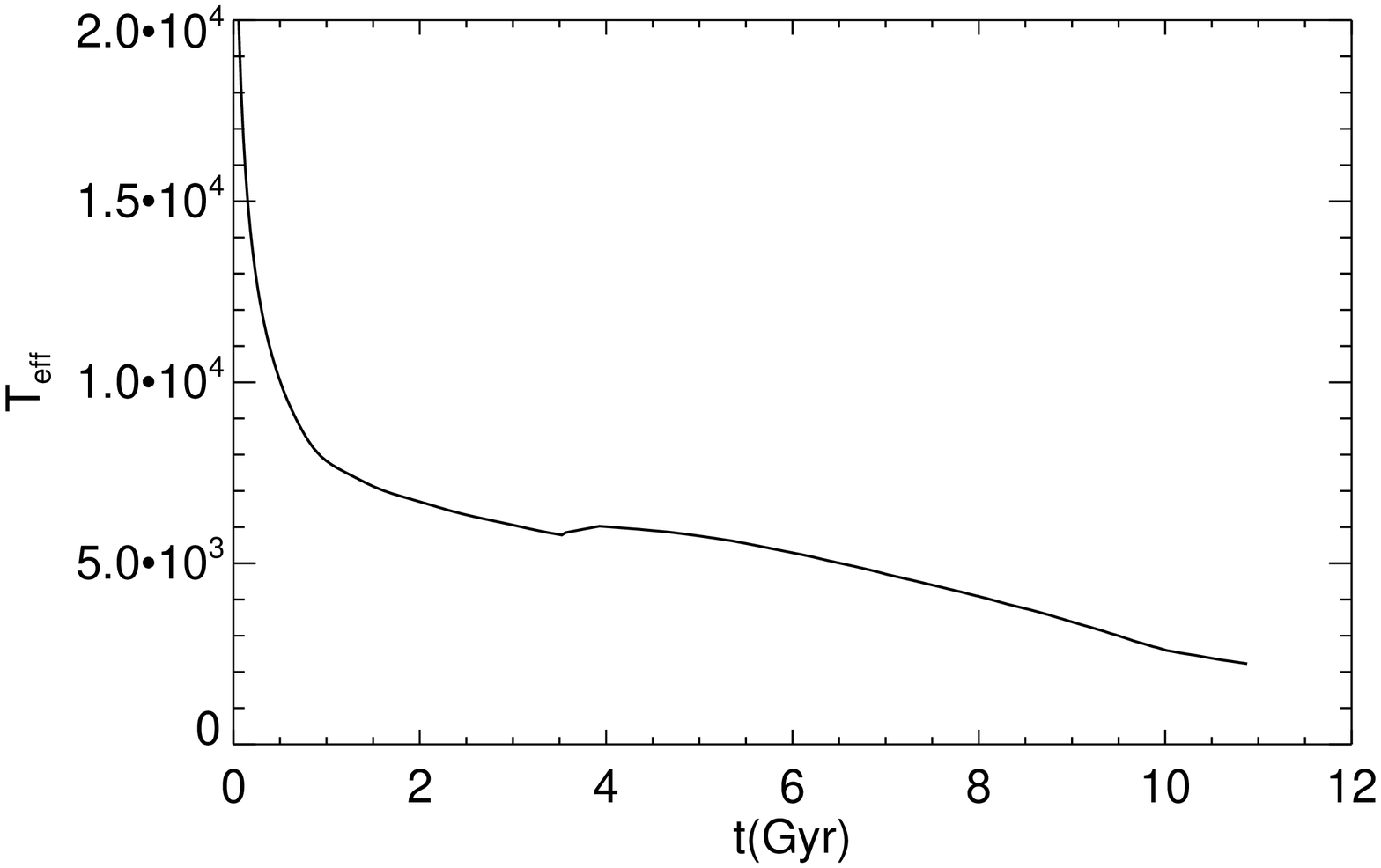}
\includegraphics[width=0.5\textwidth,height=0.3\textheight]{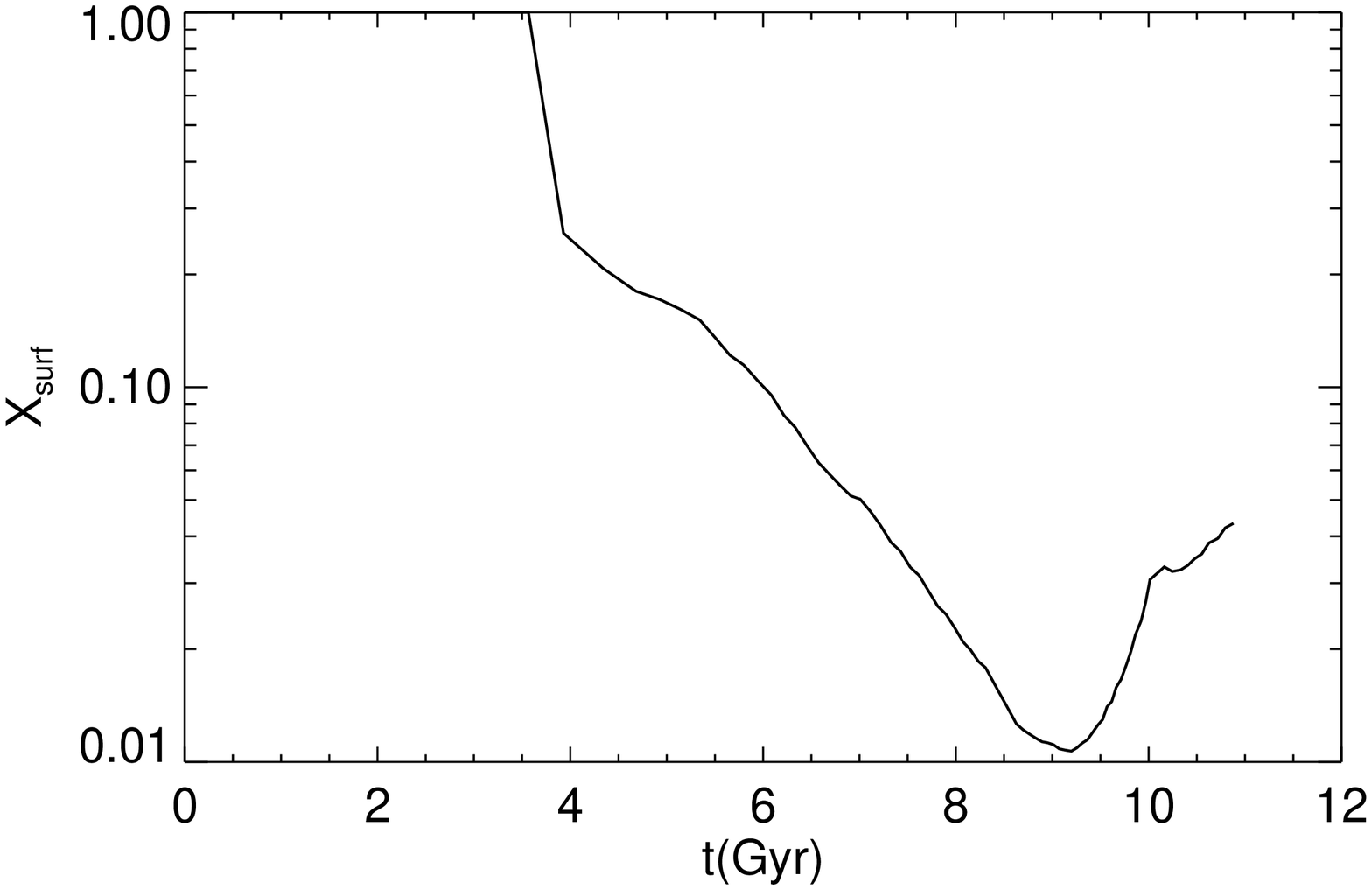}
\caption{$t_{cooling}$ vs $T_{eff}$ and $t_{cooling}$ vs $X_{surf}$ for white dwarf envelopes with $m_H=1.0\times10^{-7}M_{\sun}$.  The time spent between each white dwarf configuration is calculated by equation \ref{ttc}.  We can see here that a white dwarf of $m_H=1.0\times10^{-7}M_{\sun}$ will probably remain as a DA after dredge-up, although its $X_{surf}$ actually decreases by about two orders of magnitude.  The $T_{eff}$ does not decrease monotonically with time.  Upon dredge-up, its value increases ($\sim O(100K)$).}
\label{ploteleven}
\end{figure}

\begin{figure}
\includegraphics[width=0.5\textwidth,height=0.3\textheight]{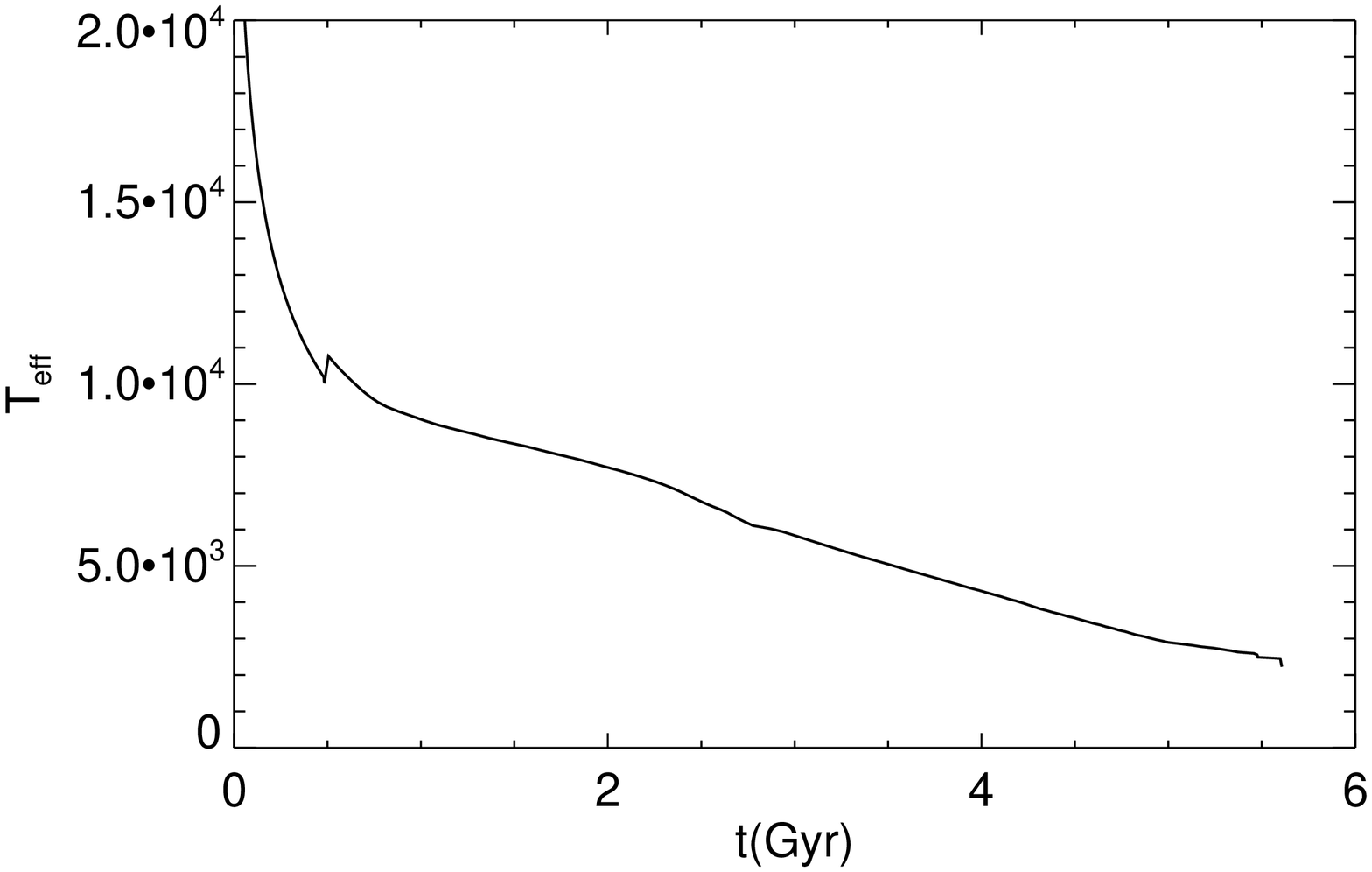}
\includegraphics[width=0.5\textwidth,height=0.3\textheight]{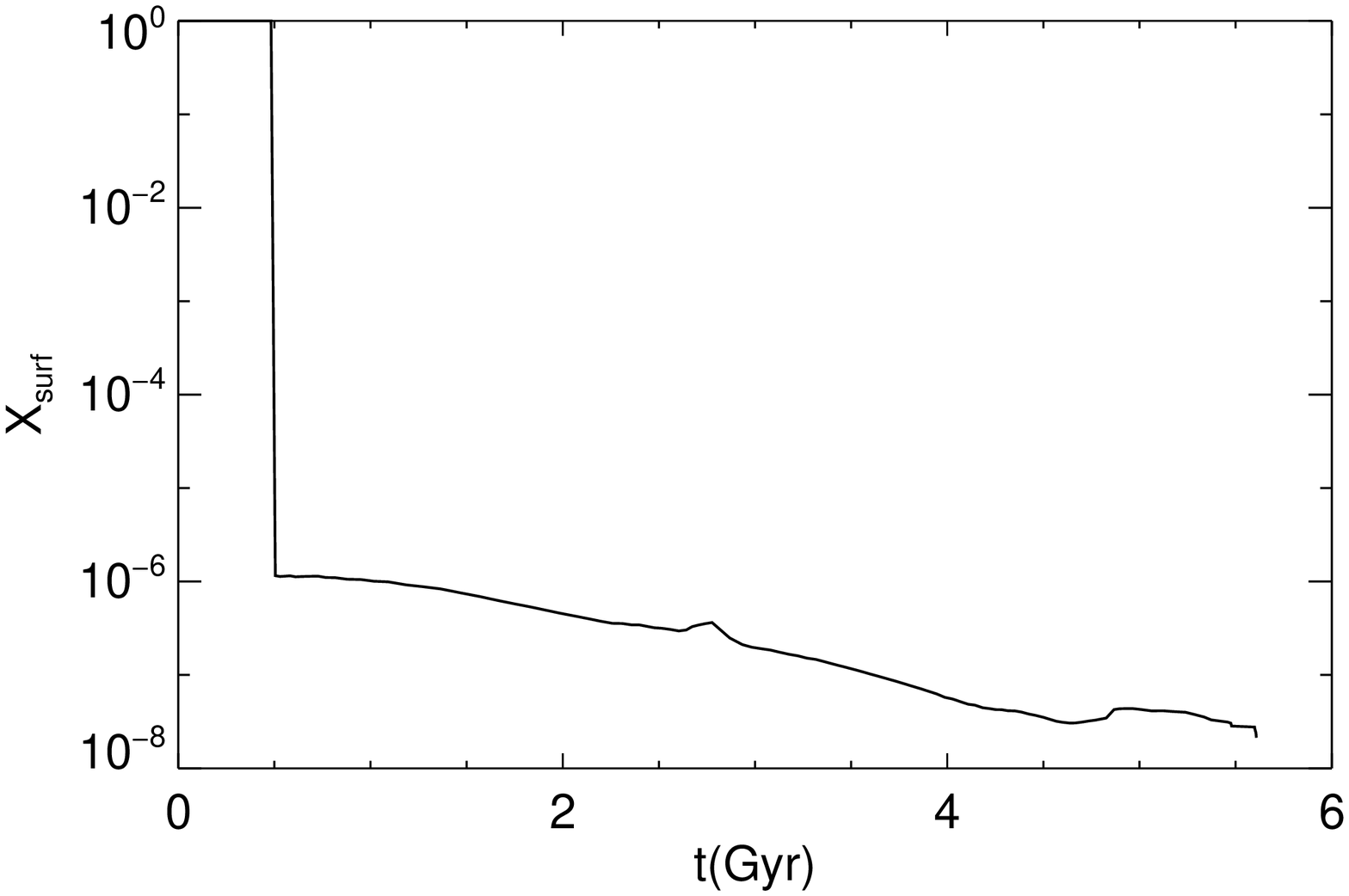}
\caption{Cooling curve and Chemical Evolution curve for white dwarf envelopes with $m_H=1.0\times10^{-11}M_{\sun}$.} 
\label{plotfifteen}
\end{figure}

A few examples of these curves are shown from figure \ref{ploteleven} to figure \ref{plotfifteen}.

\begin{figure}
\includegraphics[width=0.5\textwidth,height=0.3\textheight]{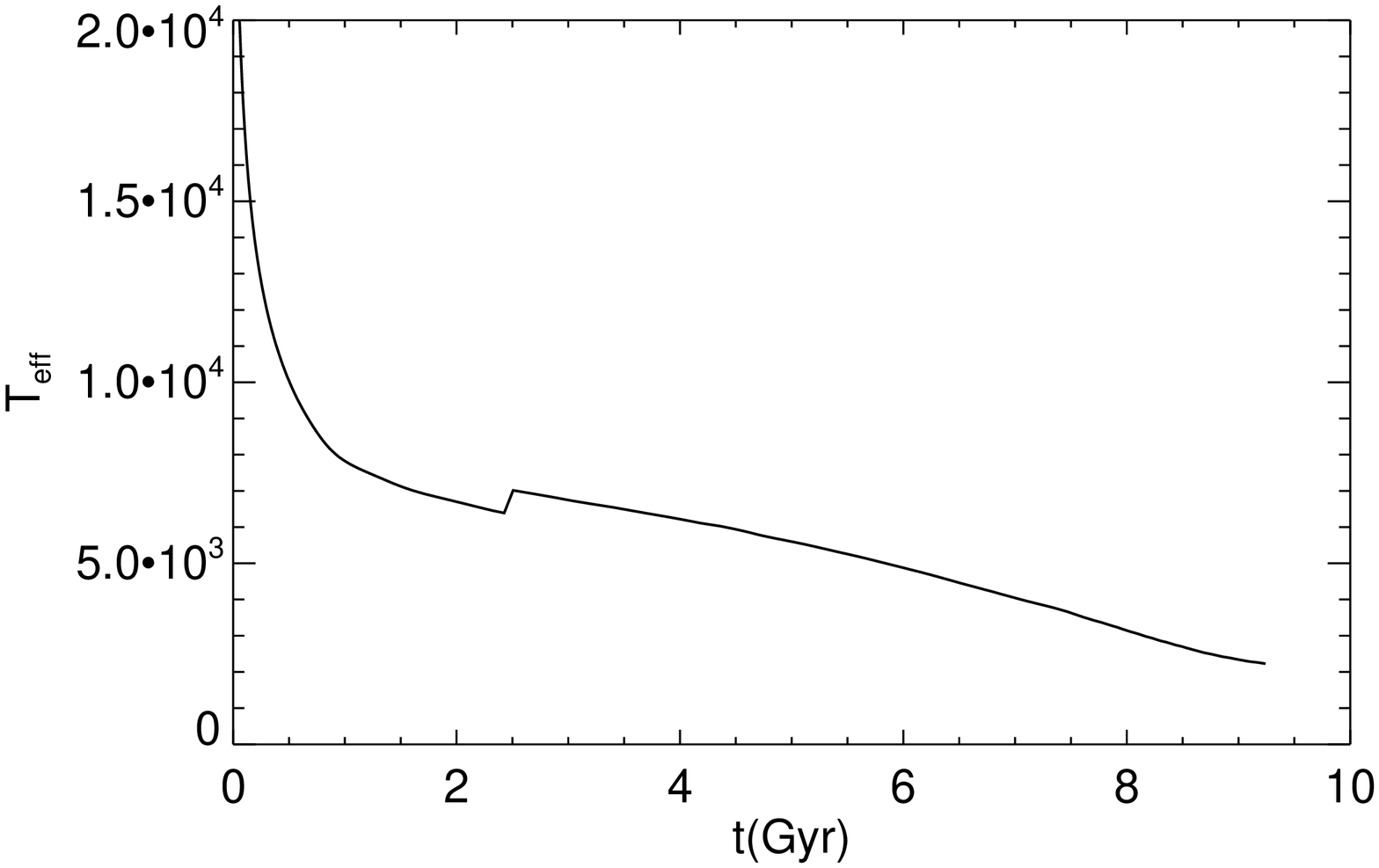}
\includegraphics[width=0.5\textwidth,height=0.3\textheight]{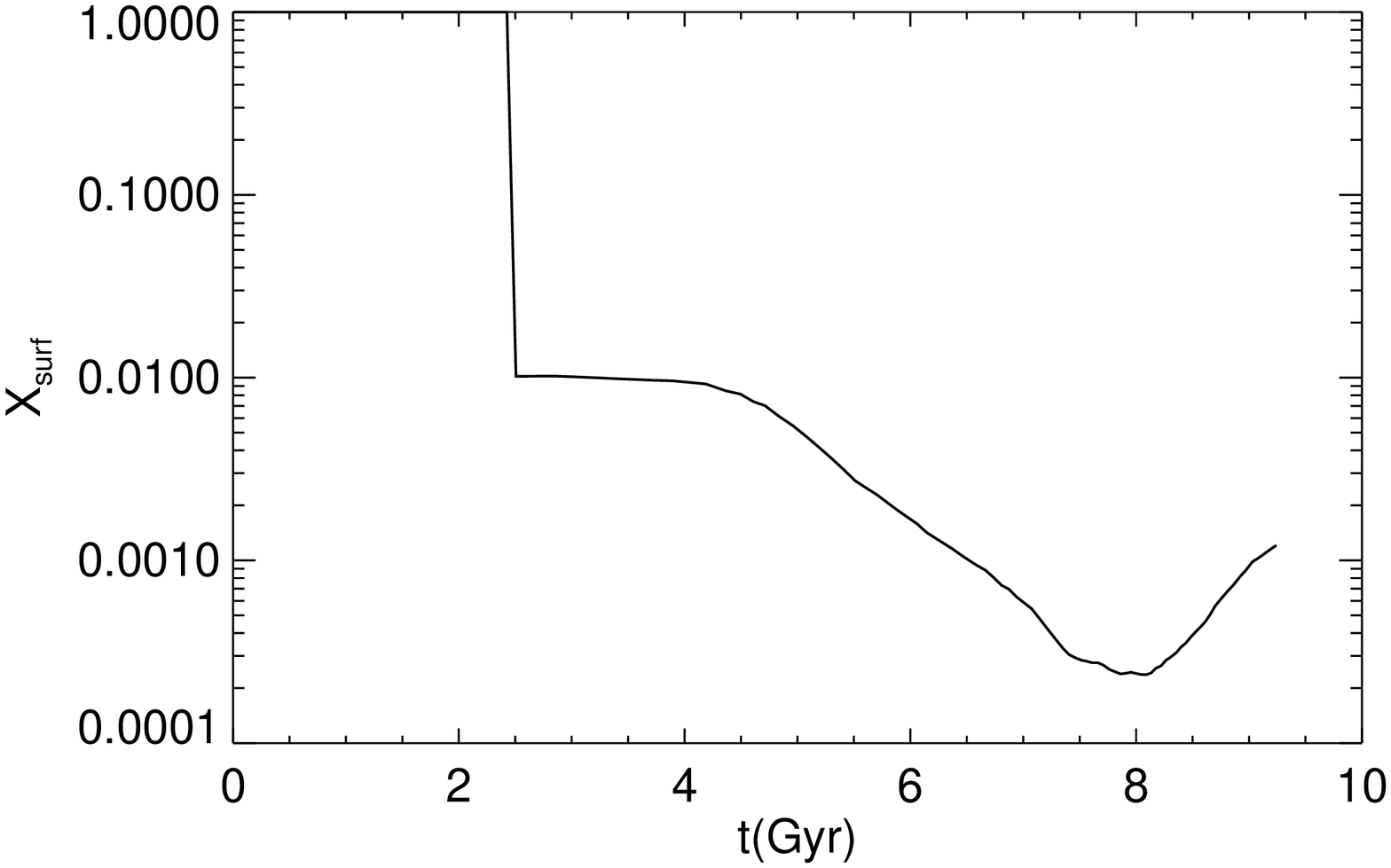}
\caption{Cooling curve and Chemical Evolution curve for white dwarf envelopes with $m_H=1.0\times10^{-8}M_{\sun}$.}
\label{plottwelve}
\end{figure}

\begin{figure}
\includegraphics[width=0.5\textwidth,height=0.3\textheight]{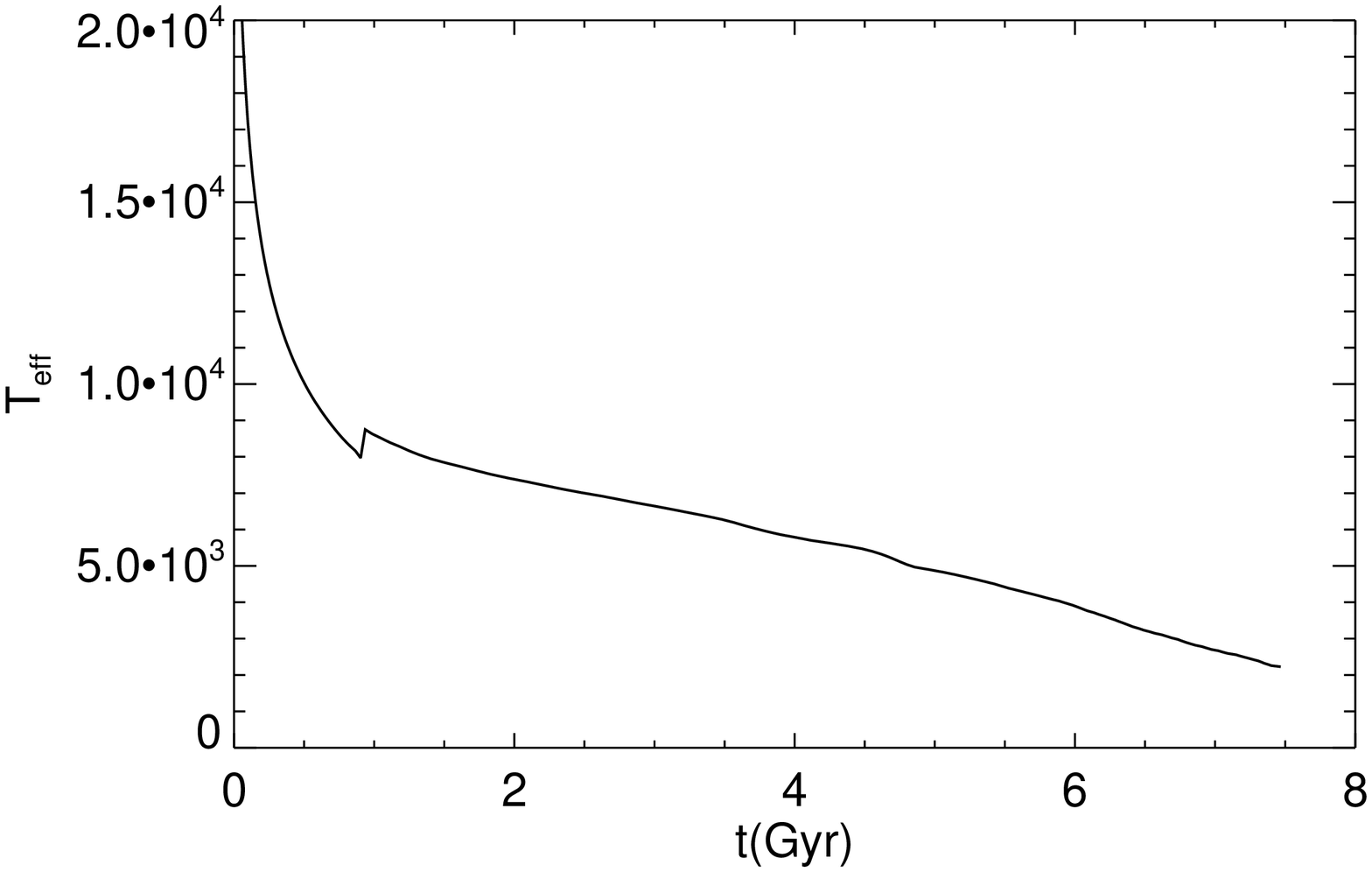}
\includegraphics[width=0.5\textwidth,height=0.3\textheight]{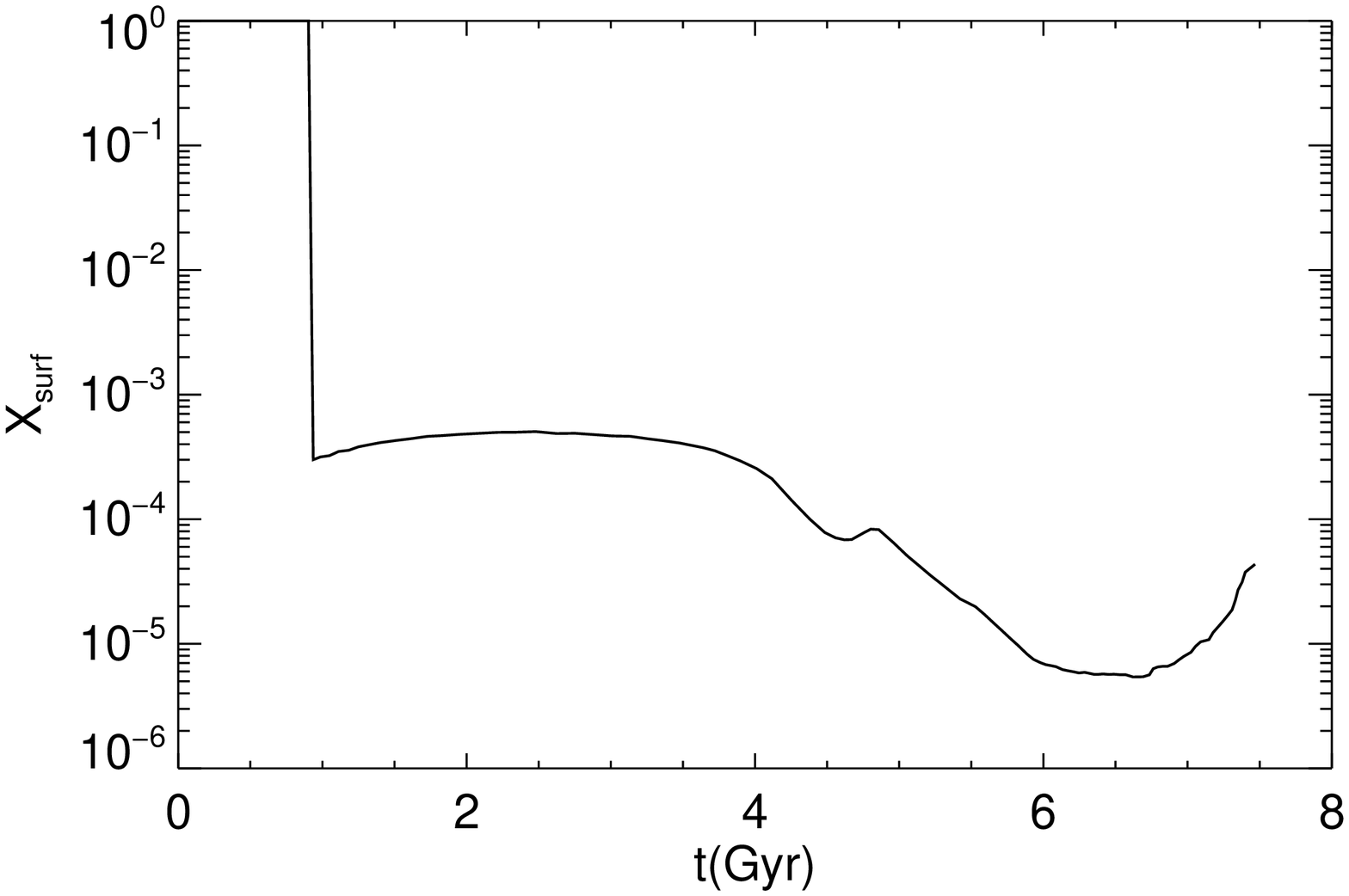}
\caption{Cooling curve and Chemical Evolution curve for white dwarf envelopes with $m_H=1.0\times10^{-9}M_{\sun}$.}
\label{plotthirteen}
\end{figure}

\begin{figure}
\includegraphics[width=0.5\textwidth,height=0.3\textheight]{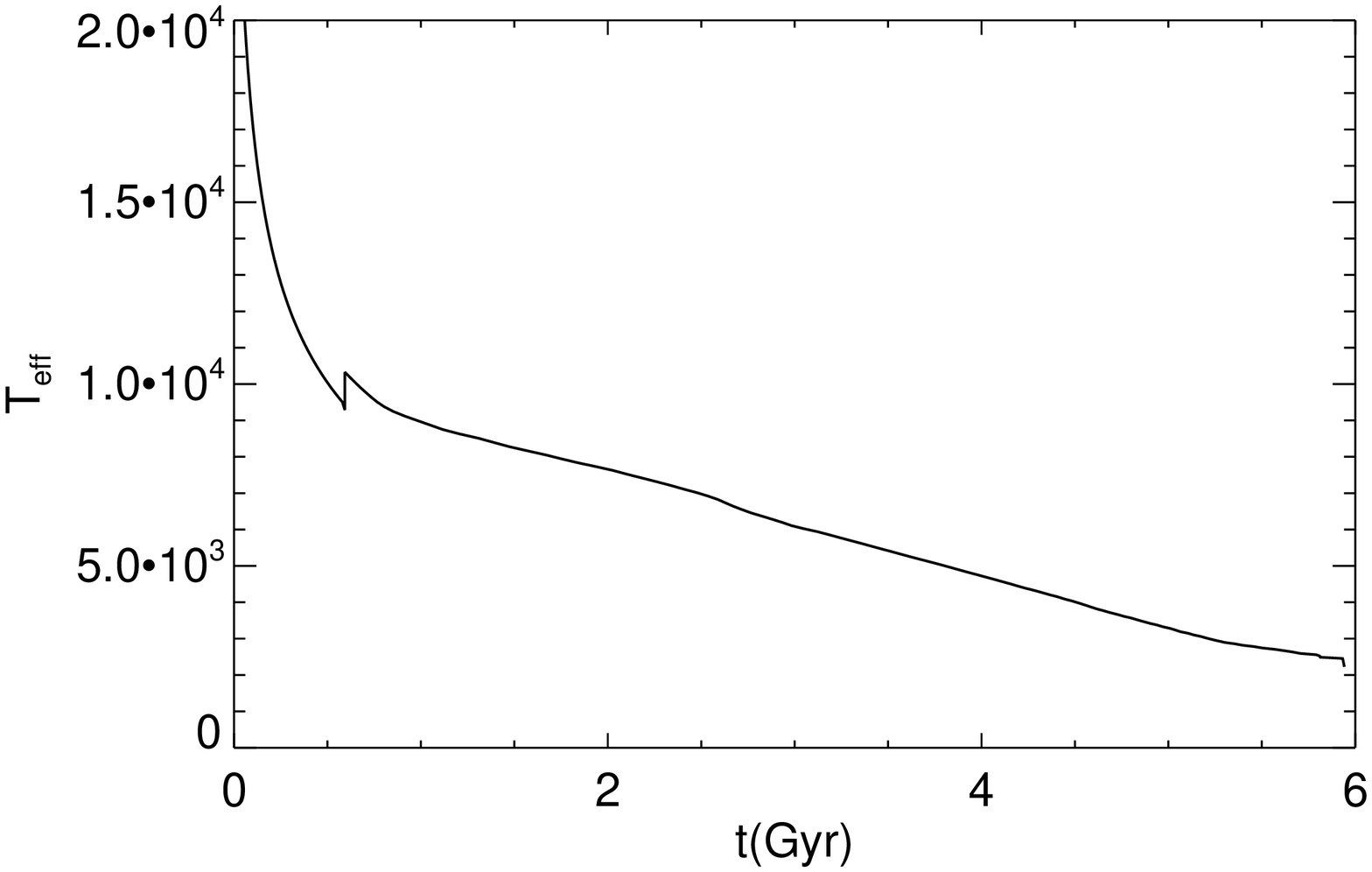}
\includegraphics[width=0.5\textwidth,height=0.3\textheight]{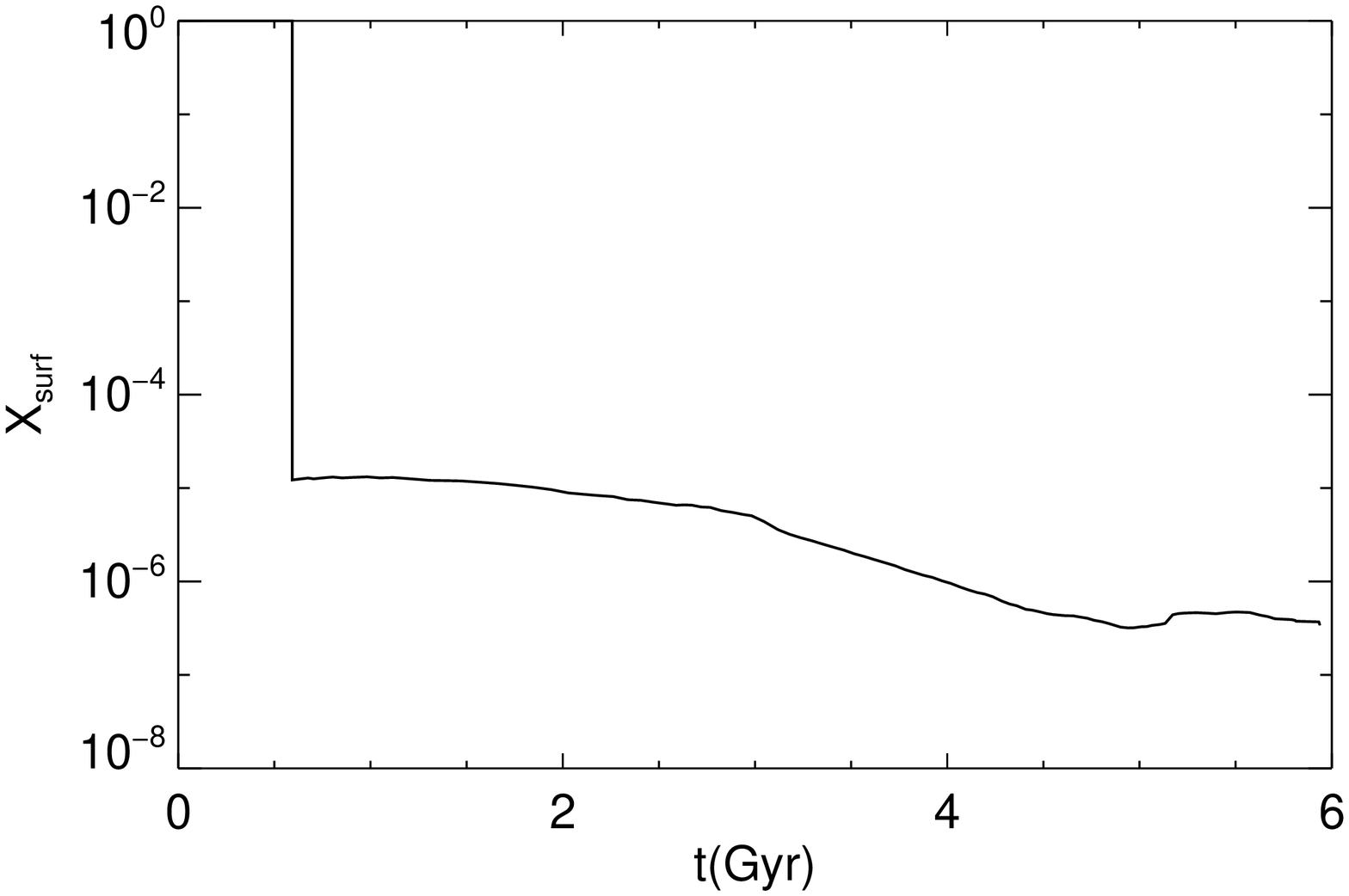}
\caption{Cooling curve and Chemical Evolution curve for white dwarf envelopes with $m_H=1.0\times10^{-10}M_{\sun}$.} 
\label{plotfourteen}
\end{figure}

\section{Discussion}
\label{sec:discuss}
In this section, we discuss the physical meanings of the cooling curves.  We also relate our results to the reported observations.

\subsection{Comparison Between Models}
We would like to compare the cooling curves for pure hydrogen stars ($m_H=10^{-5}M_\sun$), pure helium stars, and stars that undergo spectral evolution (here we use a $0.6M_{\sun}$ white dwarf with $m_H=10^{-8}M_{\sun}$ as a representative).  Strictly speaking, there are two types of cooling curves.  In the literature one usually refer ``cooling curve'' to the $t$--$T_{eff}$ curve (as we did in the previous sections).  In this section, however, we will reserve the term ``cooling curve'' for the \emph{core} cooling curve, $t$--$T_c$ relation.  $t$--$T_{eff}$ relation will be temporarily termed ``fading curve''.  This difference in terminology is made for the sake of clarity, as we will see below.

\begin{figure}
\includegraphics[width=0.5\textwidth]{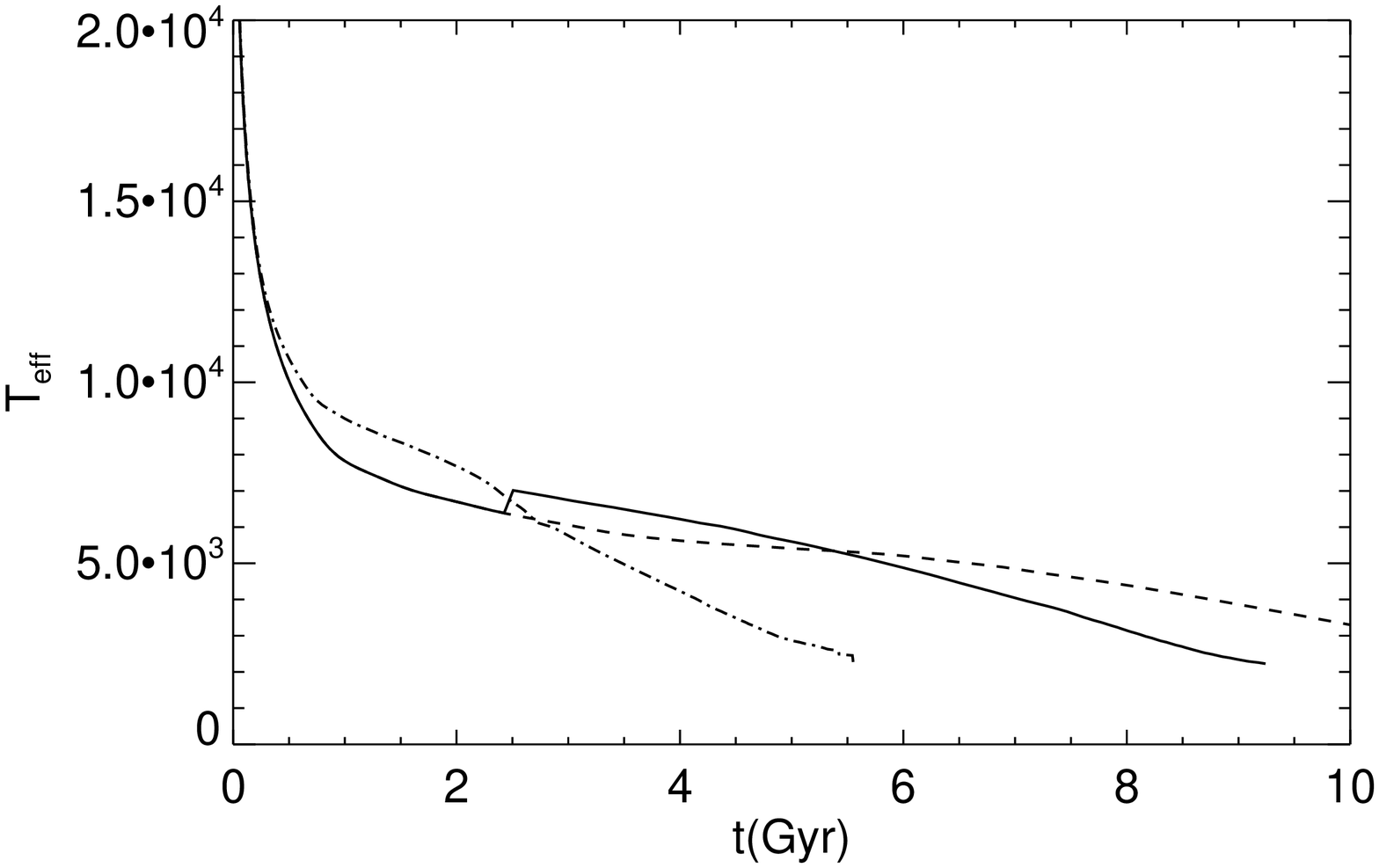}
\includegraphics[width=0.5\textwidth]{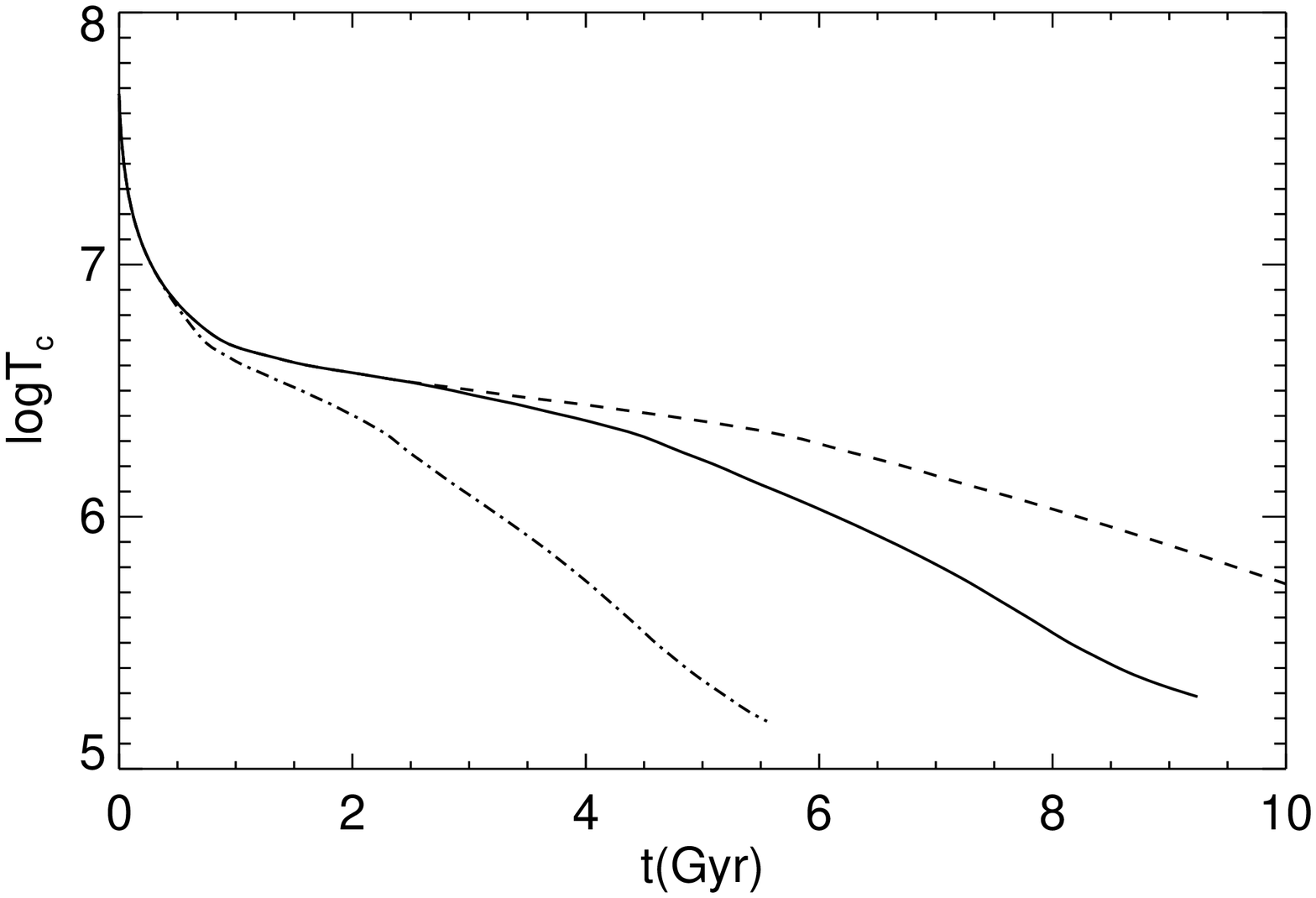}
\caption{The fading curve (upper panel) and cooling curve (lower panel) of $0.6M_{\sun}$ white dwarf with three different $m_H$:  $m_H=1.0\times10^{-5}M_{\sun}$ (dashed line), $m_H=1.0\times10^{-8}M_{\sun}$ (solid line) and $m_H=0$ (pure helium white dwarf, dash-dotted line).  The relation between cooling rate and hydrogen content is fairly simple: white dwarf with lower $X_{surf}$ cools faster.  However, the relation between fading rate and hydrogen content is more complicated and is given by equation~\ref{eqn:fadingRate}.}
\label{plotsixteen}
\end{figure}

\begin{figure}
\includegraphics[width=0.5\textwidth]{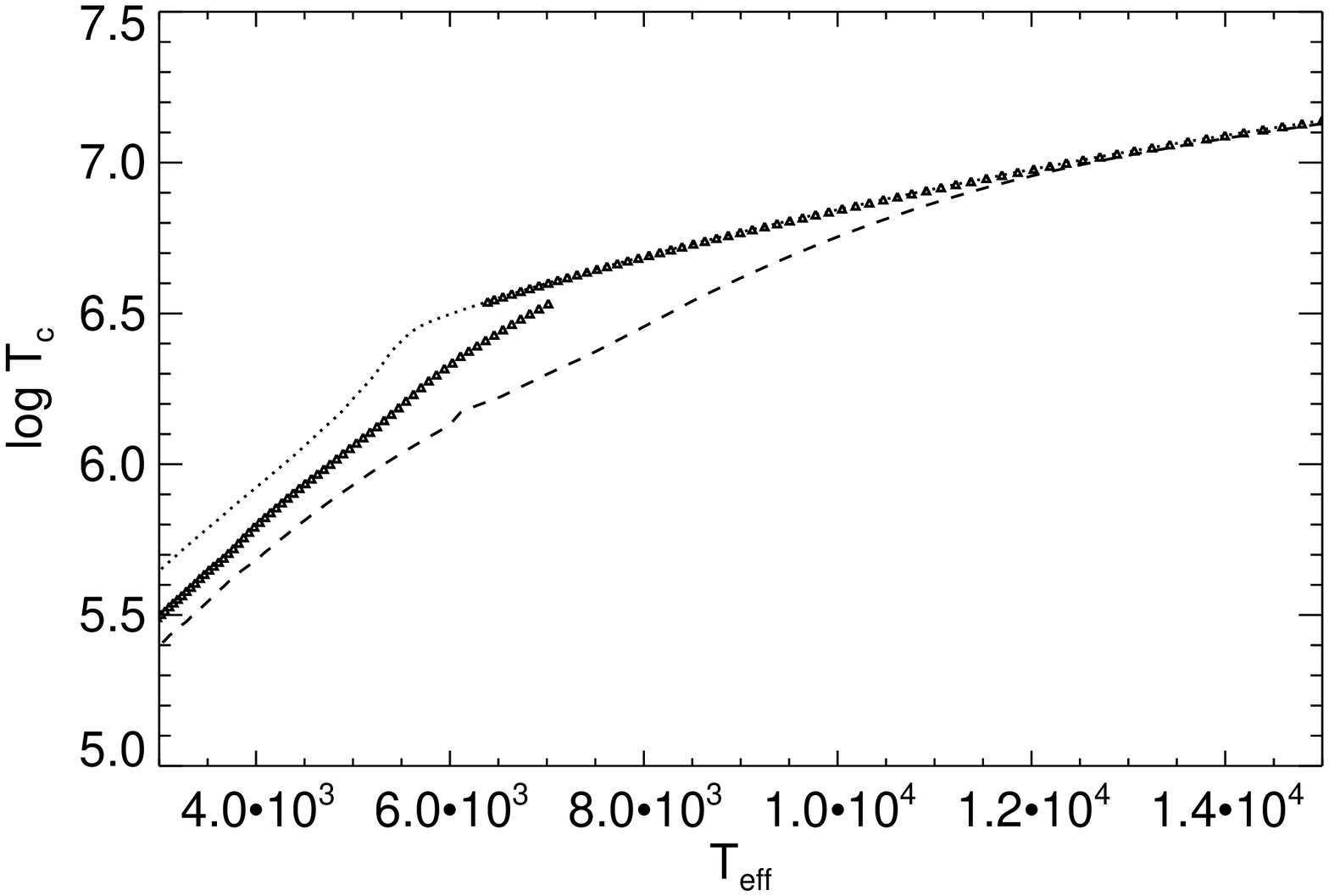}
\caption{$T_{eff}$--$T_c$ relation for white dwarf models in figure~\ref{plotsixteen}.  That of the thick hydrogen model is indicated by the dotted line and that of pure helium model is indicated by the dashed line.  The relation of the convective mixing models is represented by the triangular symbols, which lies between the two non-mixing cases.}
\label{plotseventeen}
\end{figure}

A comparison of the fading curves are shown in the upper panel of figure \ref{plotsixteen}.  The fading curve of chemical-evolving white dwarf ($m_H=10^{-8}M_\sun$) is seemingly complicated, however, in essence it is the relatively simple cooling curve (as shown in the lower panel of figure~\ref{plotsixteen}) masked by the (discontinuous) $T_{eff}$--$T_c$ relation which is given by figure \ref{plotseventeen}.  The \emph{cooling} curve in the lower panel of figure \ref{plotsixteen} shows that the cooling rate is generally higher for the white dwarfs with lower $X_{surf}$ (fixing $T_c$).

The cooling rate as a function of $T_c$ is directly determined by the relation between $T_{eff}$ and $T_c$:
\begin{eqnarray}
L=L(T_c)=-\frac{dE(T_c)}{dt}=-\frac{dT_c}{dt}\frac{dE}{dT_c}\\
\frac{dT_c}{dt}=-\frac{L(T_c)}{\frac{dE}{dT_c}}
=-\frac{4 \pi R^2 T_{eff}^4(T_c)}{\frac{dE}{dT_c}}
\end{eqnarray}
Thus, models with closer $T_{eff}$--$T_c$ relation have higher cooling rates (fixing $T_c$).  $T_{eff}$--$T_c$ relation for all three models in the convective coupling regime is shown in figure~\ref{plotseventeen}.  The triangular symbols are the relation for chemical-evolving models and the two bracketing curves are the relation for pure hydrogen stars (upper) and pure helium stars (lower).  We can see that in the regime of our interest, $T_{eff}$--$T_c$ relation is significantly different for pure hydrogen models and pure helium models.  That of chemical evolving white dwarfs, on the other hand, varies between the two extremes.  Of particular interest is the discontinuity of $T_{eff}$--$T_c$ relation located at $\log T_c=6.53$ for chemical evolving white dwarfs, which corresponds to the onset of convective mixing.  The mixing leads to a sudden increase in helium content and $T_{eff}$.

The mechanism of increase in $T_{eff}$ is a \emph{combination} of convective coupling and reduction of surface opacity.  Note that both ingredients are important.  In the complete absence of convective coupling, $T_c$ is insensitive to photosphere opacity.  It is only after convective coupling, a reduction in surface opacity allows corresponding reduction of $\log T_c$--$\log T_{eff}$ through the increment of $P_{ps}$ in equation~\ref{eq:rejuvenation}.

Fading rate, on the other hand, is determined by both cooling rate and the derivative of $T_{eff}$ to $T_c$:
\begin{equation}
\label{eqn:fadingRate}
\frac{dT_{eff}}{dt}=\frac{dT_c}{dt} \frac{dT_{eff}}{dT_c}
\end{equation}
As a result, a closer relation between $T_{eff}$ and $T_c$ does \emph{not} imply higher \emph{fading} rate.  For example, the fading rate of helium models at $t\lesssim1Gyr$ in figure~\ref{plotsixteen} is lower than that of hydrogen models, despite of their higher cooling rate.

\subsection{Comparison with Previous Works}
This work is the first study on convective mixing that accounts the post-mixing composition in a self-consistent fashion.  Besides this, however, there are additional differences between this study and previous works such as \citet{1972ApJ...177..723S}, \citet{1973A&A....27..307B} and \citet{1976A&A....52..415K}.

\citet{1972ApJ...177..723S} based his argument of $T_{eff}$ increment solely on the reduction of opacity, which is not complete.  Convective coupling is crucial for $T_{eff}$ to increase upon convective mixing.  In particular, the increment of $T_{eff}$ in the example that Shipman used ($13000K$ DA star) turns out to be negligible in our calculation due to the lack of convective coupling.

\begin{figure}
\includegraphics[width=0.5\textwidth]{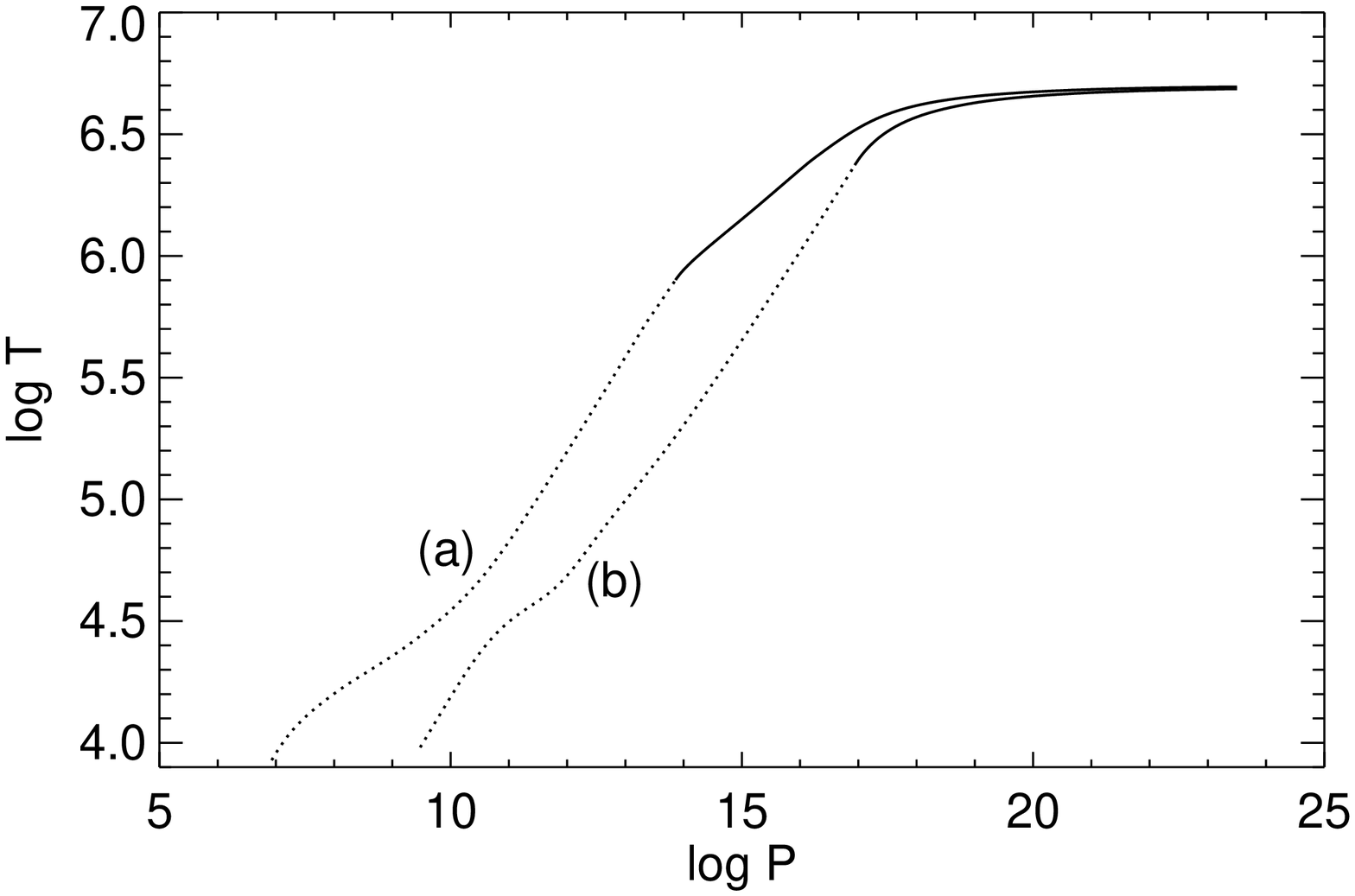}
\caption{The pre-mixing model ($X_{surf}=1$, $T_{eff}\sim7967K$, labeled by ``(a)'') and post-mixing model ($X_{surf}\sim0.0003$, $T_{eff}\sim8750K$, labeled by ``(b)'') of a $0.6M_\sun$ white dwarf with $m_H=10^{-9}M_\sun$.  We have shown that the temperature at the mass fraction which corresponds to the base of post-mixing convection zone is changed during the convective mixing process.}
\label{ploteighteen}
\end{figure}

\citet{1973A&A....27..307B} assumed the physical conditions below the post-mixing convection zone to be invariant during the mixing process, which is not accurate.  Our analysis shows a non-negligible change in $\log T$ at the mass fraction which corresponds to the base of post-mixing convection zone (c.f. figure~\ref{ploteighteen}).  We note that the luminosity is constant throughout the envelope and fixing the physical conditions at any location other than the isothermal core will naturally fix the total luminosity.  Besides, the authors claimed that the inclusion of convection zone reduces the $T_{eff}$ increment of \citet{1972ApJ...177..723S}, contrary to our findings.

The result of \citet{1976A&A....52..415K} (no luminosity change upon convective mixing) is different from us because he only investigated one particular model with extremely small $m_H$.  Due to the smallness of $m_H$, the convective mixing occurs at very high $T_{eff}$ where $T_{eff}$--$T_c$ relation is still insensitive to surface composition (see section~\ref{sec:nomix}).

\subsection{Spectral Evolution}\label{sec:spec_evol}
We have successfully calculated the evolution of $X_{surf}$ for white dwarfs with $m_H$ varying over five orders of magnitude.  However, the evolution of \emph{spectra} is not quite the same as the evolution of $X_{surf}$.  \citet{1990ApJ...351L..21B} stated that helium lines start to turn invisible at $T_{eff}$ below $\sim12000K$. As a result, white dwarf atmosphere with $\log \frac{N(He)}{N(H)}<2$ would all appear to be of spectral type DA, even if in reality, helium could be the main constituent.

In \citet{2007ASPC..372...29B}, it was again emphasized that the spectroscopic mass of low $T_{eff}$ DA white dwarfs has a much greater mean
and dispersion than its high $T_{eff}$ counterparts.  The authors thus speculated that upon convective mixing (i.e., when hydrogen convection zone breaks into the helium layer), white dwarf stars might still appear as spectral type DA instead of turning into that of spectral type DC as \citet{1972ApJ...177..723S}, \citet{1973A&A....27..307B} and \citet{1991IAUS..145..421F} have proposed.

Our results show that both proposal has its own standing.  In our higher $m_H$ models ($m_H\sim10^{-7}M_{\sun}$), the white dwarfs adjust their $X_{surf}$ to $O(0.1)$ and $T_{eff}$ increases only slightly ($\sim O(100K)$) upon convective mixing.  They will thus appear to astronomers as a slightly hotter and spectroscopically more massive DA white dwarf.

On the other hand, DA white dwarfs with $m_H<10^{-9}M_{\sun}$ will directly turn into DC stars upon convective mixing.  The $T_{eff}$ will increase at the order of $1000K$.

What we have shown is, upon convective mixing, a white dwarf star always decreases its $X_{surf}$. \emph{However, the spectral outcome and $T_{eff}$ increment will be dependent on $m_H$.}

\section{Conclusions}
\label{sec:conclude}
In this paper, we have developed a scheme to calculate the possible atmosphere configurations of a white dwarf, given its stellar mass ($M_{WD}$) and hydrogen content ($m_H$).  Based on the knowledge of these possible configurations, we worked
out the cooling curves and chemical evolution curves of a white dwarf of fixed $m_H$, undergoing convective mixing.

We would like to emphasize that although the main results are only given in a convective mixing scenario in the absence of accretion, our framework is actually versatile and is not confined to such scenario.  The result in section~\ref{sec:atmosphere} encompasses hydrogen content over multiple magnitudes and is applicable to the case where hydrogen content is variable (either due to wind or accretion).  To calculate the cooling curve in a scenario where $m_H$ is variable, one could pick the stages from the larger set of possible configurations (with many different $m_H$).  Once the configurations are picked one could calculate the cooling curve and chemical evolution curve in similar fashions.

We have adopted a semi-analytic approach to comprehensively study the possible configurations in the problem.  We have confirmed our results with full white dwarf evolutionary code (e.g., \citep{1999ApJ...520..680H}) by switching $X_{surf}$ from $1$ to the value we obtained from this work.  We found that $T_{eff}$ will automatically increase accompanying the decrease in $X_{surf}$, contrary to the results of \citet{1976A&A....52..415K}.  The flexibility
of our semi-analytic models enables us to examine the full range of behaviors associated with different hydrogen layer masses, thereby encompassing a variety of scenarios presented in the literature.

In a later paper, we will apply these cooling curves and $X_{surf}$ evolutionary curves to calculate non-DA to DA ratio as a function of $T_{eff}$, in attempt to solve the problem of the ``non-DA gap'' \citep{1997ApJS..108..339B}.  We will also obtain luminosity functions from them and discuss the impact of chemical evolution on cosmo-chronology.

\acknowledgements
The work described here is supported by NASA grant ATP03-000-0084 and the Alfred P. Sloan
Foundation.  E.C. would like to thank Dr. Bernard Freytag for discussions on the subject of convective overshooting.
\bibliographystyle{apj}
\bibliography{ms}
\newpage
\clearpage
\end{document}